\documentclass[sigconf]{acmart}

\AtBeginDocument{%
  \providecommand\BibTeX{{%
    \normalfont B\kern-0.5em{\scshape i\kern-0.25em b}\kern-0.8em\TeX}}}

\copyrightyear{2023}
\acmYear{2023}
\setcopyright{rightsretained}
\acmConference[KDD '23]{Proceedings of the 29th ACM SIGKDD Conference on Knowledge Discovery and Data Mining}{August 6--10, 2023}{Long Beach, CA, USA}
\acmBooktitle{Proceedings of the 29th ACM SIGKDD Conference on Knowledge Discovery and Data Mining (KDD '23), August 6--10, 2023, Long Beach, CA, USA}
\acmDOI{10.1145/3580305.3599470}
\acmISBN{979-8-4007-0103-0/23/08}

\usepackage{enumitem}
\usepackage{xspace}

\usepackage{soul}
\usepackage{amsmath}
\usepackage{multirow}
\usepackage{multicol}
\usepackage{tcolorbox}
\usepackage{array}
\usepackage{tabularx}
\newcommand{\model}{\textsf{INPAC}\xspace}

\newcommand{\problem}{CLIPP\xspace}
\newtheorem{prob}{\textbf{Problem}}

\begin{document}

\newcommand{\srijan}[1]{\textcolor{red}{[Srijan: #1]}}
\newcommand{\karan}[1]{\textcolor{green}{[Karan: #1]}}
\newcommand{\hide}[1]{}
\newcommand{\yc}[1]{\textcolor{blue}{YC: #1}}
\newcommand{\ks}[1]{\textcolor{magenta}{KS: #1}}
\newcommand{\yq}[1]{\textcolor{orange}{YQ: #1}}

\newcommand\red[1]{\textcolor{red}{#1}}
\newcommand{\ie}{{\textit{i.e.}}}
\newcommand{\eg}{{\textit{e.g.}}}
\newcommand{\spec}{{\it spec.}}

\title{Predicting Information Pathways Across Online Communities}



\author{Yiqiao Jin}
\orcid{0000-0002-6974-5970}
\affiliation{%
  \institution{Georgia Institute of Technology}
  \city{Atlanta}
  \state{GA}
  \country{United States}}
\email{yjin328@gatech.edu}

\author{Yeon-Chang Lee}
\orcid{0000-0002-8769-0678}
\affiliation{%
  \institution{Georgia Institute of Technology}
  \city{Atlanta}
  \state{GA}
  \country{United States}}
\email{yeonchang@gatech.edu}

\author{Kartik Sharma}
\orcid{0000-0002-0747-9320}
\affiliation{%
  \institution{Georgia Institute of Technology}
  \city{Atlanta}
  \state{GA}
  \country{United States}}
\email{ksartik@gatech.edu}

\author{Meng Ye}
\orcid{0009-0004-9290-2817}
\affiliation{%
  \institution{SRI International}
  \city{Princeton}
  \state{NJ}
  \country{United States}}
\email{meng.ye@sri.com}

\author{Karan Sikka}
\orcid{0000-0002-0187-5322}
\affiliation{%
  \institution{SRI International}
  \city{Princeton}
  \state{NJ}
  \country{United States}}
\email{karan.sikka@sri.com}

\author{Ajay Divakaran}
\orcid{0000-0003-0371-5346}
\affiliation{%
  \institution{SRI International}
  \city{Princeton}
  \state{NJ}
  \country{United States}}
\email{ajay.divakaran@sri.com}

\author{Srijan Kumar}
\orcid{0000-0002-5796-3532}
\affiliation{%
  \institution{Georgia Institute of Technology}
  \city{Atlanta}
  \state{GA}
  \country{United States}}
\email{srijan@gatech.edu}

\renewcommand{\shortauthors}{Jin et al.}

\begin{abstract}
The problem of \emph{community-level information pathway prediction} (\problem) aims at predicting the transmission trajectory of content across online communities. A successful solution to \problem holds significance as it facilitates the distribution of valuable information to a larger audience and prevents the proliferation of misinformation. 
Notably, solving \problem is non-trivial as inter-community relationships and influence are unknown, information spread is multi-modal, and new content and new communities appear over time. In this work, we address \problem by collecting large-scale, multi-modal datasets to examine the diffusion of online YouTube videos on Reddit. We analyze these datasets to construct community influence graphs (CIGs) and develop a novel dynamic graph framework, \model\ ({\bf\underline{In}}formation {\bf\underline{P}}athway {\bf\underline{A}}cross Online {\bf\underline{C}}ommunities), which incorporates CIGs to capture the temporal variability and multi-modal nature of video propagation across communities. Experimental results in both warm-start and cold-start scenarios show that \model\ outperforms seven baselines in \problem. Our code and datasets are available at \url{https://github.com/claws-lab/INPAC}
\end{abstract}





\maketitle

\section{Introduction}\label{sec:intro}


\vspace{1mm}
\noindent{\bf Background.}
Social media users form communities based on their interests, beliefs, ethnicity, and geographical location~\cite{naseem2022early, Okawa2022PredictingOD}. These communities are prevalent on popular social platforms such as Reddit, WhatsApp, and Telegram, enabling users to connect with like-minded individuals as well as consume and disseminate information in an interactive manner. As communities grow in size, they become hubs of information flow, facilitating the exchange of information across  communities. 
Existing research has shown that online communities interact with and influence one another~\cite{kumar2018community, waller2019generalists, keegan2019dynamics, fiesler2018reddit, tan2018tracing}. 


As information spreads from one community to the other, it can rapidly reach all members in the new community. 
While individual posts and hyperlinks may propagate in varying patterns, the underlying pathways on which information propagates remain relatively stable~\cite{gomez2012inferring, qiu2018deepinf}. Their stability is partially due to the behavior of common users who repeatedly spread information among the same communities, creating a reinforcing effect of the underlying information pathways. 

The fast-paced evolution of social media has accelerated the spread of information, including a variety of content types ranging from news articles, commercial advertisements, to harmful content such as online rumors~\cite{yang2022weakly, shu2019defend}, fake news~\cite{wu2022bias, yang2022reinforcement, xu2022evidence}, hate speech~\cite{Ziems2020RacismIA}, and political bias~\cite{jin2022towards}. The unmoderated spread of these contents can cause adverse social impacts. 
For example, the COVID-19 pandemic has led to the formation and growth of multiple online communities, such as subreddits \textsf{r/CoronavirusUS}, \textsf{r/COVID-19Positive}, and \textsf{r/COVID19}, where users discuss various topics related to the pandemic. These communities are inter-connected, with similar topics and user groups, thus having a significant influence on each other. Sometimes misinformation proliferates in online communities, such as the unfounded claim that 5G technology can spread the virus \cite{sturm2021constituent, ahmed2020covid}. Despite a lack of scientific evidence, this conspiracy theory gained traction in several online communities, including \textsf{r/conspiracy}, \textsf{r/5G}, \textsf{r/CoronavirusUS}, and \textsf{r/COVID19}, causing unwarranted fear and concern among the public.


\hide{The fast pace of social media has accelerated the spread of information, including a variety of content such as news articles, commercial advertisements, and even harmful content like rumors, fake news, hate speech, and political bias~\cite{shu2019defend, shu2017fake, wang2018eann}. Given these developments, predicting the transmission trajectory of information among communities on social media has become important. It requires analyzing community relationships to forecast the flow of information.
}

The Community-Level Information Pathway Prediction (\problem) problem seeks to predict the transmission trajectory of information among online communities. 
\problem is of significant importance as it enables prediction of communities where information, including problematic content, is likely to emerge and spread. Such capability can provide numerous benefits across a wide range of applications. Efficient prediction of misinformation spread with \problem can guide intervention strategies, while for advertising, \problem can refine strategies and maximize the efficacy of marketing campaigns, increasing the visibility of information and providing insights into the communities where their target audience is most active. 


\hide{
Users organize themselves into communities on web platforms.
User-defined communities are an essential component of many
web platforms, where users express their ideas, opinions, and share members of one community and engage with members of another. Studies of intercommunity dynamics in the offline setting have shown
information. These communities not only provide a gathering place
for intracommunity interactions between members of the same
community, they also facilitate intercommunity interactions, where
These communities can interact with one another, often leading to
conflicts and toxic interactions. However, little is known about the
mechanisms of interactions between communities and how they
impact users.
However, analyses of intercommunity interaction and conflict
are largely absent in previous work on web communities.
}



\hide{
For this reason, there has been a marked increase in research initiatives aimed at comprehending the information spread on various social media platforms from different perspectives, such as the impact of misinformation~\cite{jin2022towards, yang2022reinforcement, yang2022weakly, xu2022evidence, wu2022bias}, the effect of social media on individual and collective behavior~\cite{shareef2020group, zhou2012social}, and the analysis of information diffusion from an algorithmic perspective~\cite{myers2012information, yang2010modeling}.  
Among these, the problem of {\bf Community-level Information Pathway Prediction (\problem)} has gained particular attention in recent years due to its wide-ranging practical applications~\cite{li2021capturing, feng2022h, gomez2013structure}. 
}

 
\hide{Furthermore, recent works have focused on investigating the information propagation between individual users, \ie, ``microscopic'' influence, and predicting information pathways on the user-user level~\cite{chen2022ctl, zhou2020variational, li2017deepcas, myers2014bursty}. 
}


\vspace{1mm}
\noindent{\bf Challenges.}
Solving \problem is challenging. First, community-to-community influence is usually unknown~\cite{gomez2012inferring, rong2016model}, and the
mechanisms of interactions between communities and how they
impact users remains hidden~\cite{kumar2018community}. 
Different communities may have different norms, values, and communication patterns that influence the temporal patterns of information diffusion~\cite{xia2021deepis}. In this case, we only observe where new content is propagated to the new communities and when it takes place. 
The underlying community influence, \ie, who influences the propagation, remains unknown. Most existing works focus on predicting information diffusion at the user level (\ie, microscopic influence)~\cite{qiu2018deepinf, leung2019personalized}.
Meanwhile, existing datasets~\cite{verma2022examining, Micallef2022CrossPlatformMM, Ziems2020RacismIA, Micallef2020TheRO} only contain limited information about community structures, making it difficult to study cross-community information spread. 

Second, the spread of information is characterized by a complex and dynamic diffusion environment \cite{lin2017better}. Posts contain multi-modal signals, such as text, images, and videos~\cite{chen2022cross, java2009we, chang2016positive}. 
Diffusion patterns vary across content types. 
For example, misleading news and inflammatory microblogs spread faster and wider than true information~\cite{vosoughi2018spread,jenders2013analyzing,gustafsson2010time}. 
Niche content are usually shared within a few narrow-interest communities, while broad-interest contents create far-reaching cascades and reach several disparate communities~\cite{waller2019generalists, waller2021quantifying, phadke2021makes}. 
Understanding these propagation patterns is essential to predicting information spread across communities.

\hide{
Effective solutions to the IPP problem can bring significant benefits to various organizations and individuals.
In online marketing, businesses can launch more successful marketing campaigns by 
identifying key influencers [cite]. 
In combating misinformation, online moderators can better identify vulnerable individuals susceptible to certain false information~\cite{gomez2013structure, ng2022cross}. 
}

\vspace{1mm}
\noindent{\bf Our Work.}
In this work, we investigate the dynamics of community-level information flow while jointly addressing the challenges of complex diffusion environment and the continuously evolving information ecosystem. 

We choose Reddit as the platform for studying community-level information diffusion since it provides numerous communities, named ``subreddits,'' that are dedicated to specific topics or interests. 
Towards this goal, we collect two large-scale and multi-modal datasets that enable us to study the community-level diffusion of visual contents for information pathway prediction. Based on that, we identify distinct temporal patterns of information sharing using inter-activity time distribution, infer macroscopic community-to-community influence, and construct novel community influence graphs (CIGs). 

We design \model, or {\bf\underline{In}}formation {\bf\underline{P}}athway {\bf\underline{A}}cross Online {\bf\underline{C}}omm unities, a dynamic graph-based method to predict community-level information pathways using CIGs and content's multi-modal information (visual features and channel metadata). 
\model integrates structure, content semantics, and temporal information by utilizing Continuous-Time Dynamic Graphs (CTDGs) to model the time-aware propagation patterns of videos. In \model, nodes and edges are continuously introduced to the graph, incorporating both visual features and channel metadata of the content. 

\hide{
Then, we design \model , or {\bf\underline{In}}formation {\bf\underline{P}}athway {\bf\underline{A}}cross Online {\bf\underline{C}}ommunities, while modeling cross-community information flow using our datasets (\eg, Table~\ref{tab:subredditsSharingVideo} \srijan{why is this table referenced here?}). 
Specifically, we leverage the structural information as manifested in the content sharing events of online communities. 
By doing so, we can capture correlations between visual contents and communities and address the rapid shifts of information diffusion, thereby successfully addressing (C2). 

Moreover, \model\ is based on the observation that users exhibit similar temporal patterns in their content sharing behavior. This allows us to utilize the time intervals between consecutive shares as a powerful indicator of influence. This simple yet powerful method enables us to characterize the relationships of influence among communities and infer the most likely communities to share a video with limited information.
Thus, we provide excellent inductive capabilities on unknown contents (\ie, (C3)).
}





\vspace{1mm}
\noindent{\bf Contributions.}
Our main contributions are as follows:
\begin{itemize}[leftmargin=*]
    
    \item \textbf{Novel Multi-modal Datasets and Analysis}: 
    We collect two large-scale, multi-modal datasets to study community-level diffusion of visual contents for information pathway prediction. 
    We identify distinct temporal content sharing patterns that are used to infer community-to-community influence graphs. 

    \item \textbf{Information Pathway Prediction Framework}: To solve CLIPP, we propose \model, a dynamic graph framework based on CIGs that learns from multimodal data and the dynamics of the interactions between users and communities. 
    
    \item \textbf{Experimental Evaluation}: We demonstrate the effectiveness of \model framework and its design choices through experiments in various scenarios, \eg, prediction of cold/warm-start videos on communities with various popularity. \model reaches performance improvements of up to 18.8\% on MRR, 13.8\% on NDCG@5, and 6.2\% on Rec@5.
\end{itemize}

\begin{table}[t]
\vspace{-0.1cm}
\caption{Statistics of our datasets. 
}
\vspace{-0.2cm}
\label{tab:data}
\centering
\begin{tabular}{c|cccc} 
\toprule
 & \textsf{Large} & \textsf{Small} \\ 
\midrule
\#Videos / URLs & 183,596 & 6,802 \\
\#Subreddits & 57,894 & 7,319 \\
\#Users & 291,047 & 8,752 \\ \midrule
\#Shares & 1,323,714 & 36,118 \\
Density & 7.96E-05 & 6.11E-04 \\ 
\midrule
\#Cold-start Videos & 3,042,068 & 68,095 \\
\bottomrule 
\end{tabular}
\vspace{-0.3cm}
\end{table}

\section{Dataset and Problem}

\subsection{Dataset Description}\label{sec:data}
In this study, we aim to study the spread of visual content across communities on social media.
To this end, we collect massive visual contents on YouTube and long-term community activity on Reddit.
The reasons for selecting these two platforms in this study as follows: 
\begin{itemize}[leftmargin=*]
    \item \textbf{YouTube} is one of the most widely used video-sharing platforms that contains over 2.56 billion users\footnote{\url{https://www.statista.com/statistics/272014/global-social-networks-ranked-by-number-of-users/}} and provides a venue for users to upload, share, and view videos.
    \item \textbf{Reddit} is one of the largest social platforms for content creation, rating, and sharing. It allows users to interact in a variety of communities (\ie, subreddits). Reddit is an ideal platform for studying the propagation of online visual contents such as YouTube videos because of its vast and diverse user base as well as its open-source nature and community structures. 
\end{itemize}


\begin{table*}
\small
\centering
\caption{Examples of cross-community information flow in our datasets. A video is usually shared on a set of semantically similar subreddits. ``$\rightarrow$'' indicates the temporal order of the sharing. 
}
\vspace{-0.2cm}
\label{tab:subredditsSharingVideo}
\begin{tabular}{c|c} 
\toprule
\toprule
\textbf{Title of the Video} & \textbf{Subreddits on Which the Video Appears}  \\ 
\midrule
Canadian Trudeau Investigation & \begin{tabular}[c]{@{}c@{}}Liberate\_Canada $\rightarrow$ conspiracy $\rightarrow$ TheNewRight $\rightarrow$ \\PeoplesPartyofCanada $\rightarrow$ Canada\_First\end{tabular} \\ 
\hline
Reviews: Super Dragon Ball Heroes Episode 19 & \begin{tabular}[c]{@{}c@{}}promote $\rightarrow$ AnimeReviews $\rightarrow$ anime\_manga $\rightarrow$\\YouTubeAnimeCommunity $\rightarrow$ Anime\_and\_Manga\end{tabular} \\ 
\hline
Warcraft 3 Reforged Cutscene Only & WC3 $\rightarrow$ pcgaming $\rightarrow$ warcraft3 $\rightarrow$ gaming $\rightarrow$ legaladviceofftopic \\ 
\hline
Practical Greeting Phrases for Chinese New Year & learnchinese $\rightarrow$ learnmandarin $\rightarrow$ learnmandarinchinese \\ 
\hline
Accepting what is. (Realize Instant Freedom) & \begin{tabular}[c]{@{}c@{}}AnxietyDepression $\rightarrow$ Soulnexus $\rightarrow$~SpiritualAwakening $\rightarrow$\\Meditation $\rightarrow$ spirituality $\rightarrow$ awakened $\rightarrow$ inspiration\end{tabular} \\ 
\hline
Covid-19 Explained with Data Science & \begin{tabular}[c]{@{}c@{}}Python $\rightarrow$ CoronavirusUS $\rightarrow$ CanadaCoronavirus $\rightarrow$\\CoronaVirus\_2019\_nCoV $\rightarrow$ CoronavirusUK\end{tabular} \\ 
\hline
\multicolumn{1}{l|}{Implement RNN-LSTM for Music Genre Classification} & learnmachinelearning $\rightarrow$ Python $\rightarrow$ tensorflow $\rightarrow$ musictheory \\
\bottomrule
\bottomrule
\end{tabular}
\vspace{-0.2cm}
\end{table*}

As the first step, we collected 54 months of historical Reddit posts from January 2018 to June 2022 via \textsf{PushShift}\footnote{\url{https://pushshift.io/}}. 
We removed any posts that did not contain valid URLs 
and retained URLs associated with valid YouTube videos, resulting in 5,723,910 posts and 3,737,191 associated videos. Finally, following previous works~\cite{harper2015movielens, chen2019joint}, we retained videos shared within at least 3 communities. 
Table~\ref{tab:data} shows the statistics of the two datasets we construct.
The \textsf{large} dataset covers 54 months of video propagation history from January 2018 to June 2022, while the \textsf{small} dataset covers a 3-month period from January to March 2020. Table~\ref{tab:data} reveals that both datasets contain a considerable number of cold-start videos with only one interaction in a subreddit, which reflects the real-world distribution and the challenges associated with information pathway prediction.

\subsection{Problem Formulation}
We formulate the \problem problem as follows: 
Given a video and a sequence of subreddits in which it has been posted, predict the next community the video will be posted in at a given time. 
Formally, we define a posting of a video as a video link appearing on a subreddit, either as a standalone post or as part of a longer post. 
A \emph{posting instance} is represented as a 4-tuple $p_{ij} = (v_i, s_j, u_j, t_j)$, where $v_i$ is a video posted 
by a user $u_j$ in an online community $s_j$ at time $t_j$. 
The \emph{posting sequence} for $v_i$ is defined as a list of posting instances $P_i = \{(v_i, s_j, u_j, t_j)\}_{j=1}^N$, which indicates the dissemination trajectory with length $N$ across communities for the video $v_i$. Then, our problem can be defined as follows:
\begin{tcolorbox}
\begin{prob}[\textbf{Information Pathway Prediction}]\label{prob:ipp}
Given a video $v_i$, its posting sequence $P_i = \{(v_i, s_j, u_j, t_j)\}_{j=1}^N$ with length $N$, and a target timestamp $t_{j'}$, our model outputs a ranked list of communities $\{s_k\}$ indicating the most likely communities that $v_i$ will appear at time $t_{j'}$.
\end{prob}
\end{tcolorbox}

\begin{table}
\small
\centering
\vspace{-0.1cm}
\caption{Notations used in this paper.}
\vspace{-0.2cm}
\label{tab:notations}
\begin{tabular}{c|l} 
\toprule
\textbf{Notation} & \textbf{Description} \\ 
\midrule
$V, S$ & Set of videos and communities \\
$v_i, s_j, u_k$ & Video, community, user \\
$S^u$ & Historical interaction sequence for user $u$ \\ \midrule
$P_i$ & Posting sequence of video $v_i$ \\
$\mathcal{G}_i^S$ & Community-community influence graph for $v_i$ \\
$\mathcal{G}^D$ & Dynamic graph \\
$n$ & Maximum sequence length \\
$e_{jk}$ & Edge weights \\ \midrule
$\alpha$ & Teleport probability for APPNP \\
$\lambda_1, \lambda_2$ & Hyperparameters \\
$\Delta_t^{\text{Same}}, \Delta_t^{\text{Diff}}$ & Time intervals for same / different users \\
$f_{\theta}(\cdot, \cdot)$ & Message function for dynamic modeling \\
\bottomrule
\end{tabular}
\vspace{-0.2cm}
\end{table}

Table~\ref{tab:notations} 
summarizes a list of notations used in this paper.


\section{The Proposed Framework: \model}

\begin{figure*}
    \centering
    \includegraphics[width=0.96\textwidth]{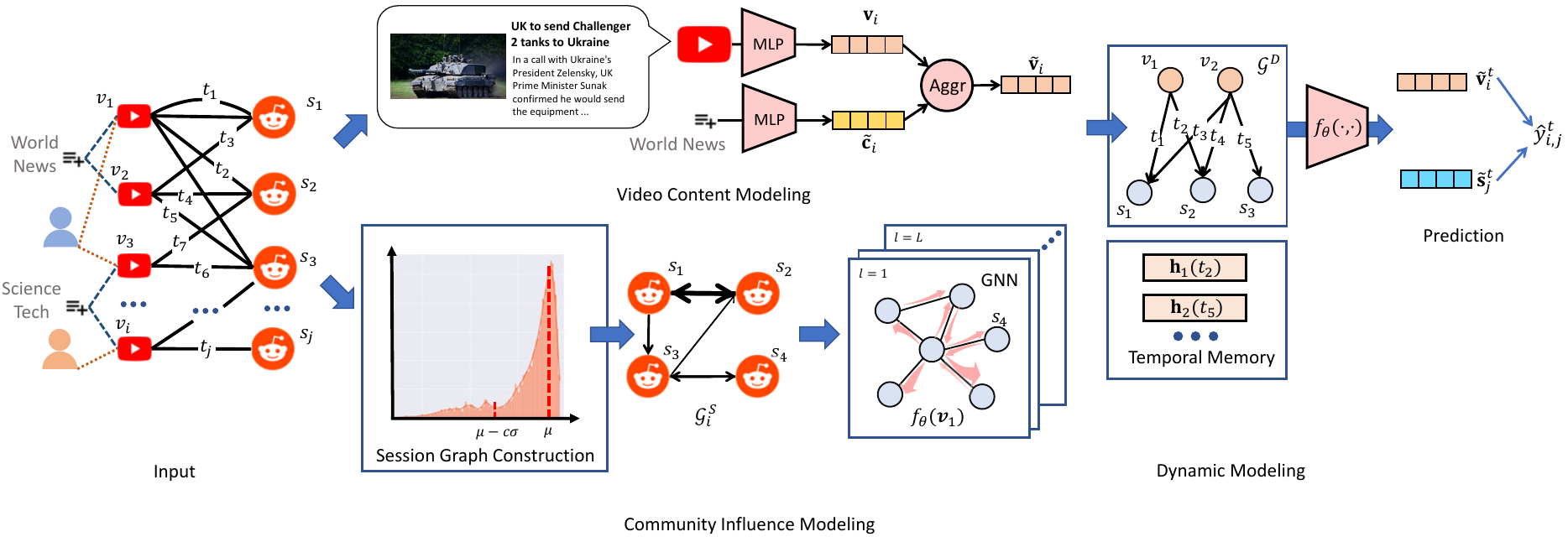}
    \caption{The overview of our proposed \model framework, which consists of static modeling, including video content and community influence modeling, as well as dynamic modeling.
     }
    \label{fig:model}
\end{figure*}

\subsection{Overview}
In this work, we aim to study the propagation of online visual content on social media.
To this end, we propose a dynamic graph framework \model\ based on Community Influence Graphs (CIGs) that learns the dynamics of cross-community information flow and accurately predicts information pathways.
As shown in Figure~\ref{fig:model}, \model\ consists of three key modules: (1) community influence modeling; (2) video content modeling; and (3) dynamic modeling.

\subsection{Community Influence Modeling}
\label{sec:community}

Given a community (\eg, a subreddit), \model\ learns its embedding such that the embedding preserves its influence on other communities during information propagation. 
We infer the influence relationships between communities using content sharing patterns in those communities. 
Specifically, a video is usually shared in communities that have similar topics. 
For example, in Table~\ref{tab:subredditsSharingVideo}, the video ``Practical Greeting Phrases for Chinese New Year'' is shared within a set of subreddits related to language learning, such as \texttt{r/learnchinese} and \texttt{r/learnmandarin}. 
To model this, we create a novel influence network by leveraging the video's temporal interaction patterns. 
\hide{Therefore, we first construct the network over which information propagates between communities.
Then, we capture the structural dependency in the reconstructed network, obtaining each community's feature vector.}


\vspace{1mm}
\noindent \textbf{Influence Graph Construction}. 
In the context of \problem, community-level influence is defined as the presence of causal relationships between posting of a video in two different communities. 
This can happen when two communities share a common group of users. To infer the influence exerted by one community on another, we employ a sequence of communities $\{s_1, s_2, \ldots\}$, in which a video $v_i$ is posted. 
\begin{figure}[h]
    \centering
    \vspace{-0.2cm}
    \includegraphics[width=0.47\textwidth]{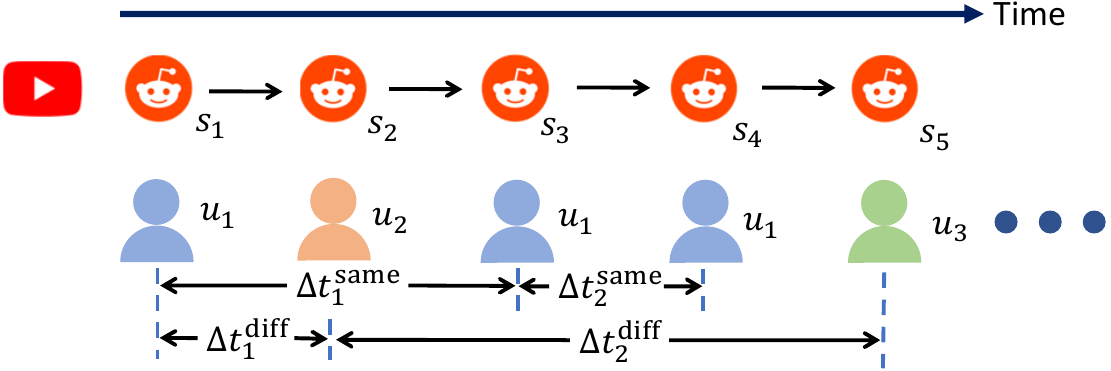}
    \caption{Illustration of how $\Delta t^{\text{Same}}$ and $\Delta t^{\text{Diff}}$ are calculated. 
    }
    \vspace{-0.3cm}
    \label{fig:how_to_construct_delta_t}
    \vspace{-0.5cm}
\end{figure}
Assuming that users require a certain amount of time to engage in online content, the interval between the appearance of a video $v_i$ in two communities $s_1$ and $s_2$ serves as an indicator of the influence of $s_1$ on the appearance of the video in $s_2$. 
If a video is shared by two users within a very short time interval,
it suggests that the shares occur simultaneously and are not influenced by one another. 
Based on this assumption, we model a posting sequence $P_i$ of $v_i$ among communities as a directed graph $\mathcal{G}_i^{S}$ consisting of community nodes $s_j$ involved in the propagation of a video $v_i$.

To model the propagation sequence of a video, we first identify its concurrent sharing events, 
where the propagation of the video takes place within a brief time period, referred to as a session, in the same or different communities. 
To this end, one needs to decide whether two shares are within the same session. 
A straightforward approach is to set a threshold time limit, such as one hour or one day, as is common in session-based recommender systems~\cite{wu2019session, chen2020handling, guo2019streaming, li2017neural, ludewig2018evaluation} 
However, this ad-hoc use of the time limit is insufficient as it can vary across datasets, videos, and platforms~\cite{oh2022implicit, jannach2017recurrent}. 

We note that consecutive sharing of a video can occur due to the same user or different users, 
resulting in differing sharing patterns and motivations. 
Therefore, we create two distributions of time difference between consecutive shares of each video $v_i$: (1) $\Delta t^{\text{Same}}$, representing the time intervals between consecutive shares of $v_i$ by the \emph{same user}; 
(2) $\Delta t^{\text{Diff}}$, representing the time intervals between the first share of $v_i$ by \emph{different users}.
Figure~\ref{fig:how_to_construct_delta_t} illustrates an example of a video's consecutive sharing on several communities over time by three users.
From Figure~\ref{fig:how_to_construct_delta_t}, we can observe how the two time intervals $\Delta t_{1}^{\text{same}}, \Delta t_{2}^{\text{same}}$ for the same user $u_1$ as well as the two intervals $\Delta t_{1}^{\text{diff}}, \Delta t_{2}^{\text{diff}}$ for different users $u_1, u_2, u_3$ are computed.


For $\Delta t^{\text{Same}}$, 
it is important to consider that a user's multiple postings of the same video in different communities should not be viewed as one community influencing another. 
This is because users usually post the same content in various venues to enhance its visibility and attract more ``likes"~\cite{xu2012modeling, firdaus2018retweet, jiang2020clicking}. 
This is not indicative of natural flow of content from one community to another. 

\begin{figure}
    \includegraphics[width=0.23\textwidth]{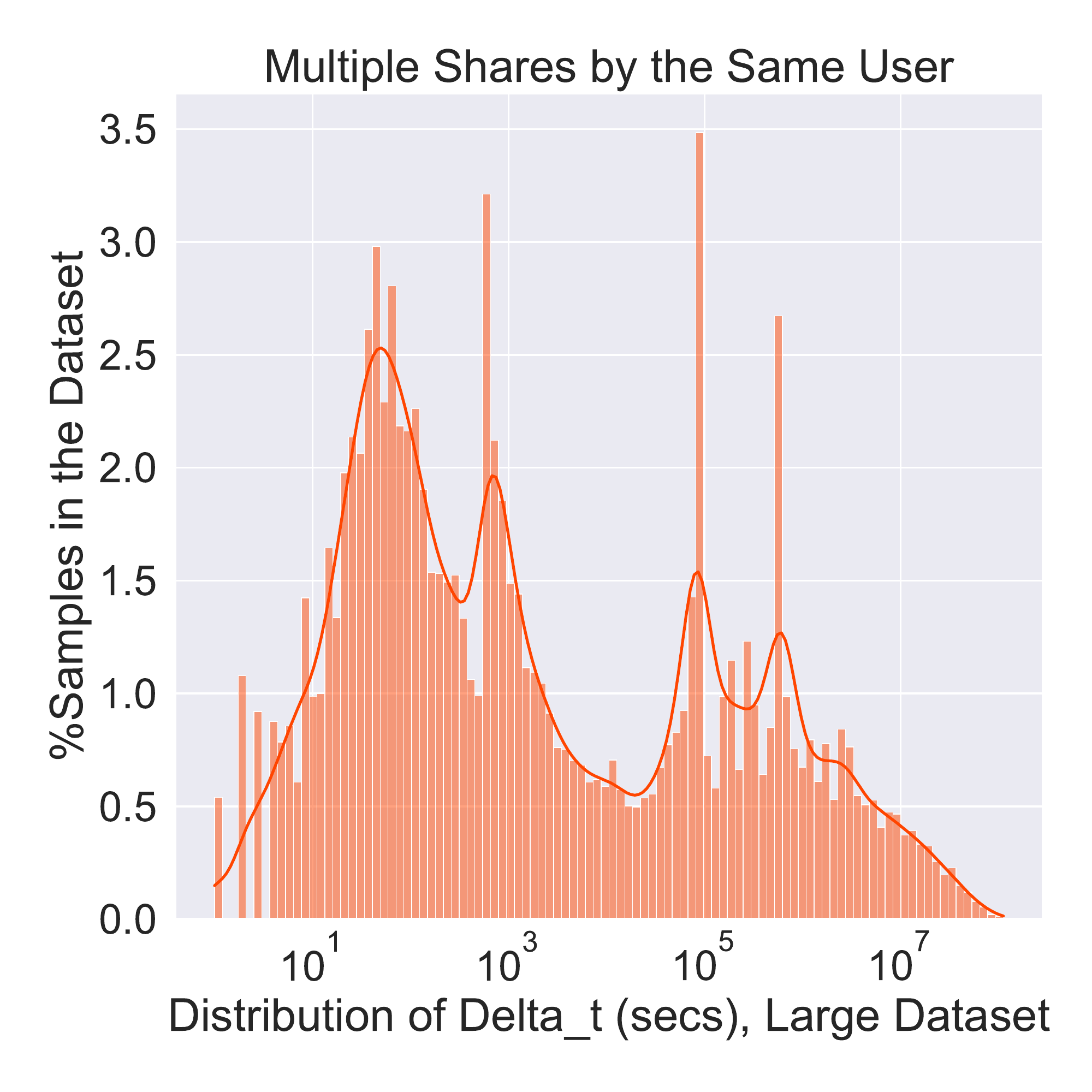}
    \includegraphics[width=0.23\textwidth]{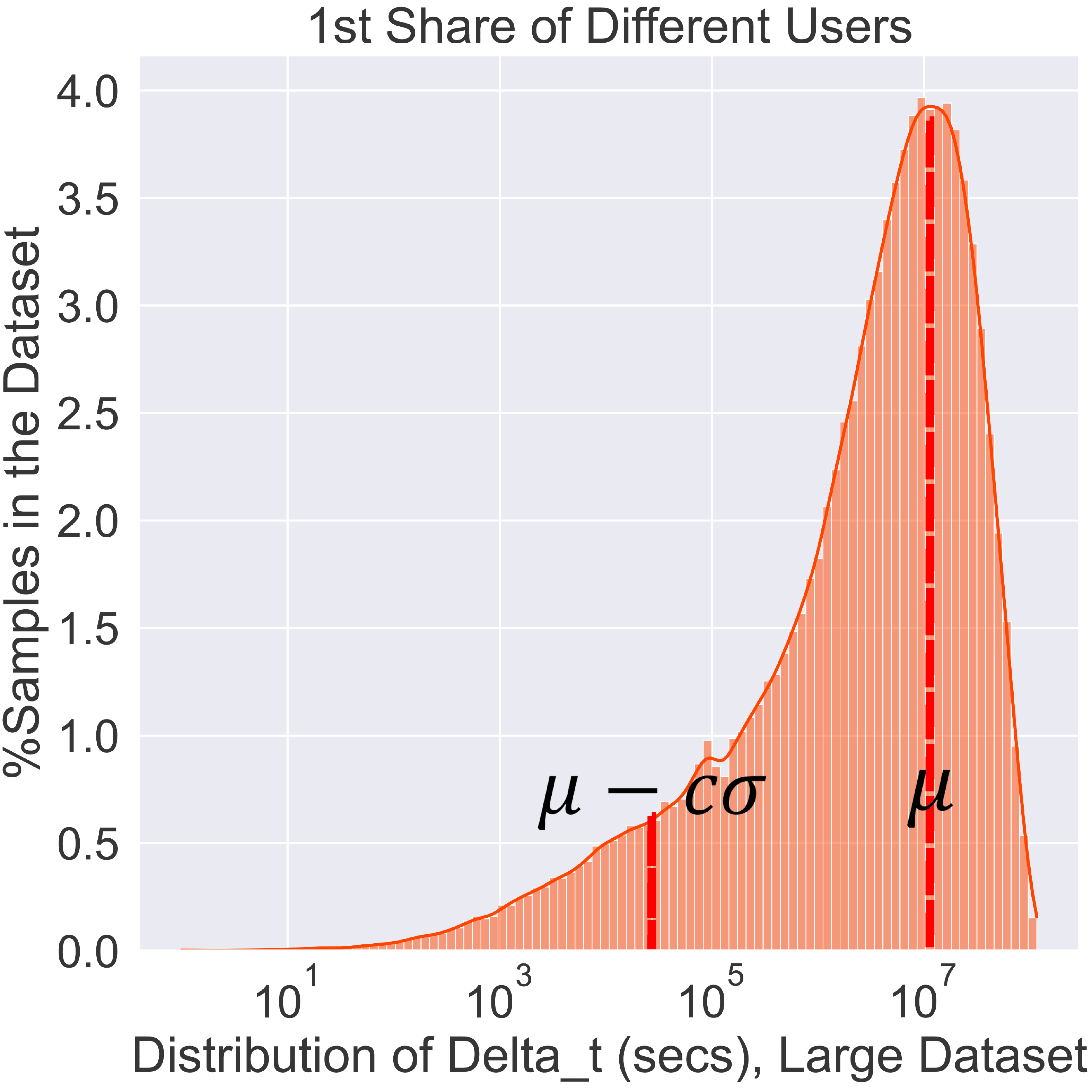}
    \caption{Distribution of $\Delta t^{\text{Same}}$ (Left) and $\Delta t^{\text{Diff}}$ (Right) for videos on the Large dataset. 
    }
    \label{fig:deltaTLargeDiff}    
\end{figure}

Thus, we only utilize $\Delta t^{\text{Diff}}$ to identify community-level influence. 
Specifically, we plot the distribution of $\Delta t^{\text{Diff}}$ across sharing events of all videos, as shown in Figure~\ref{fig:deltaTLargeDiff}, where the $x$-axis represents the time interval in seconds with a logarithmic scale of base 10, and the $y$-axis indicates the percentage.
Then, we fit a Gaussian distribution to $\Delta t^{\text{Diff}}$ and found that the distribution has a mean of 6.844 
and a standard deviation of 0.823 on the logarithmic scale. 
Based on this finding, we determine the cutoff time for partitioning sessions using $\Delta t^{\text{Thres}} = \mu-c\sigma$, where $c$ is a hyperparameter that represents the confidence level for determining concurrent shares. 
When the time difference between two postings exceeds $\Delta t^{\text{Thres}}$, the later posting is considered to be influenced by the former. 


Now, we construct the community influence graph (CIG) $\mathcal{G}_i^{S}$ with respect to $v_i$ based on the threshold $\Delta t^{\text{Thres}}$. 
Each node in $\mathcal{G}_i^{S}$ indicates a community $s_j$ and a directed edge from $s_j$ to $s_k$ indicates $s_k$ is influenced by $s_j$.
Specifically, if two shares of $v_i$ from different users occur within $\Delta t^{\text{Thres}}$, they are considered concurrent postings in the same session and not influenced by each other. 
Otherwise, a directed edge is added from $s_j$ to $s_k$ for $t_j < t_k$ in $\mathcal{G}_i^{S}$. 
Furthermore, when $v_i$ is simultaneously shared by the same user in two different communities $s_j$ and $s_k$, a bi-directional edge is added between these communities to reflect their mutual influence as a result of overlapping users.



\vspace{1mm}
\noindent \textbf{Message Aggregation}.
After the construction of $\mathcal{G}_i^{S}$, the graph is transformed from a multigraph to a weighted graph by merging multiple edges with the same source and destination nodes. 
Let $\mathcal{E}_{jk}$ denote the set of edges between $s_j$ and $s_k$. The new edge weight $e_{jk}$ is calculated using the logarithmic value
\begin{align}
    e_{jk} = \ln(1 + |\mathcal{E}_{jk}|).
\end{align}

As $\mathcal{G}_i^{S}$ consists of a number of periphery nodes, such as inactive online communities with few propagations, long-range dependencies should be considered to learn distinct node representation. 
To this end, we leverage the propagation scheme of APPNP~\cite{gasteiger2018predict} based on the personalized PageRank algorithm~\cite{brin1998anatomy}. 
APPNP adds a probability of teleporting back to the root node, which ensures that the PageRank score encodes the local neighborhood for every node and mitigates the oversmoothing issues. 

Then, we obtain the embedding matrix $\mathbf{S}_i^{(l)}$ at layer $l$ for communities involved in the $i$-th propagation sequence $P_i$:
\begin{align}
\mathbf{S}_i^{(l)} & =(1-\alpha) \hat{\mathbf{D}}_i^{-1 / 2} \hat{\mathbf{A}}_i \hat{\mathbf{D}}_i^{-1 / 2} \mathbf{S}_i^{(l-1)}+\alpha \mathbf{S}_i^{(0)},
\label{eq:appnp}
\end{align}
where $\mathbf{S}_i^{(0)} = [\mathbf{s}_1 || \ldots || \mathbf{s}_{|\mathcal{G}_i^{S}|}]$ is the initial embedding matrix for all $s_i \in \mathcal{G}_i^S$. $\hat{\mathbf{A}}_i$ and $\hat{\mathbf{D}}_i$ are the adjacency matrix and the diagonal degree matrix, respectively. $\alpha \in [0, 1)$ is the teleport probability. 

During training, we derive a probability distribution over all communities $\mathbb{P}(s_{N+1} | v_i, \mathcal{G}_i^{S})$, which indicates the most likely community for the next share of $v_i$. This requires both the current status of the sharing and the global information about $\mathcal{G}_i^{S}$. The current status can be represented using the latest posting event encoded in $\mathbf{s}_{|\mathcal{G}_i^{S}|}$. For global information, we leverage soft-attention to derive $\beta_i$, the importance of each community in the posting sequence 
\begin{align}
\beta_i &=\mathbf{w}_1^{\top} \sigma\left(\operatorname{Linear} \left(\mathbf{s}_{|\mathcal{G}_i^{S}|} \right) + (\operatorname{Linear} \left(\mathbf{s}_i \right)\right),
\end{align}

\noindent where $\mathbf{w}_1 \in \mathbb{R}^d$ is trainable parameter. $\sigma(\cdot)$ is the sigmoid activation function.

Finally, we compute the probability by taking linear transformation over the concatenation: 
\begin{equation}
\mathbb{P}(s_{N+1} | v_i, \mathcal{G}_i^{S})=\operatorname{Softmax}(\operatorname{MLP}(\left[\mathbf{s}_{|\mathcal{G}_i^{S}|} || \sum_{i=1}^n \beta_i \mathbf{s}_i\right])
 \mathbf{S}_i
),
\end{equation}
where $||$ is the concatenation operand. $\mathbf{S}_i = \left[\mathbf{s}_1 || \mathbf{s}_2 || \ldots || \mathbf{s}_{|\mathcal{G}_i^{S}|}\right]$ is the concatenation of all community embeddings in the sessions.

\subsection{Video Content Modeling}\label{sec:video}

Given a video, \model\ encodes its visual content into a low-dimensional feature vector.
The content modeling component of \model can utilize a diverse range of encoders. Here, we note that online visual content is highly diverse in terms of topics, languages, and subject matter. Therefore, the titles, descriptions, and metadata of these videos such as channel information, can provide valuable insights into their content. These additional data sources can be leveraged to better categorize and understand the content of videos. 
We thus utilize the titles, descriptions, and channel information as the static features for each video. 
Specifically, inspired by the success of pre-trained language models in natural language understanding~\cite{devlin2019bert, jin2022prototypical, wang2023cat, wang2023car}, we encode the title and descriptions of each video $v_i$ into a feature vector $\mathbf{v}_i \in \mathbb{R}^{D}$ based on a multilingual version of MiniLM~\cite{wang2020minilm}. 
Similarly, we encode each video's channel $c_{\rho(i)}$ into a feature vector $\mathbf{c}_{\rho(i)}$, where $\rho(\cdot): V \rightarrow C$ maps each video to the channel that posts $v_i$. 
Then, the two feature vectors are aggregated into a joint representation 
\begin{align}
    \label{eq:AggregateEmbeddings}
    \mathbf{\tilde{v}}_i=\mathrm{Aggr}(\mathbf{v}_i, \mathbf{c}_{\rho(i)}).
\end{align}

Here, a wide variety of aggregation schemes can be applied, including addition, concatenation, and element-wise multiplication, to obtain the joint representation. 
In Section~\ref{sec:ablation}, we investigate the impact of using different aggregation schemes for $\mathbf{v}_i$ and $\mathbf{c}_{\rho(i)}$. 

\subsection{Dynamic Modeling}
\label{sec:dynamic}

In the dynamic modeling component, \model\ models the temporal variability of each video's propagation on communities, obtaining temporal embedding of videos and communities. 
Here, we note that a video can be shared multiple times within a short amount of time~\cite{li2013popularity}. Inspired by continuous-time dynamic graph (CTDG)~\cite{rossi2020temporal, zhang2023continual, TGC_ML, S2T_ML}, we design a dynamic modeling module to provide a robust representation of the video sharing process and better handles the bursty nature of information sharing.

First, we leverage temporal graph network (TGN)~\cite{rossi2020temporal} and represent our dynamic network as a pair $(\mathcal{G}_0^{D}, E)$ where $\mathcal{G}_0^{D}$ is the initial state of the dynamic network represented as a static graph. $E$ is a set of graph events with timestamps. 
In \model, we consider two types of graph events, including node additions (\ie, the emergence of new videos and communities) and edge additions (\ie, a video is posted in an online community). 

\vspace{1mm}
\noindent \textbf{Input Encoding}. 
The input embeddings $\mathbf{x}_i(t)$ and $\mathbf{x}_j(t)$ are raw feature representations for each video $v_i$ and community $s_j$, respectively. We leverage the embeddings derived from Section~\ref{sec:community}-\ref{sec:video} 
as the raw node embeddings. Namely, $\mathbf{x}_i(t) = \mathbf{\tilde{v}}_i$ for video $v_i$ and $\mathbf{x}_j(t) = \mathbf{s}_j^{(L)}$ for community $s_j$, where $\mathbf{s}_j^{(L)}$ is the representation of $s_j$ at the final layer in Equation~\ref{eq:appnp}. 

\vspace{1mm}
\noindent \textbf{Time Encoding}.
Similar to~\cite{rossi2020temporal, xu2020inductive, song2021temporally}, the time encoding function $\phi(\cdot): \mathbb{R} \rightarrow \mathbb{R}^{d}$ maps a continuous timestamp to the $d$-dimensional vector space: 
\begin{equation}
\phi(t)=\cos \left(t \mathbf{w}_2+\mathbf{b}_1\right),
\end{equation}
where $\mathbf{w}_2, \mathbf{b}_1 \in \mathbb{R}^{d}$ are learnable parameters. 

\vspace{1mm}
\noindent \textbf{Temporal Memory}.
As in~\cite{rossi2020temporal}, to track the propagation state for each node, $v_i$ or $s_j$, at any timestamp, 
there exists a memory vector, $\mathbf{h}_i(t)$ or $\mathbf{h}_j(t)$, to store history interactive memory in a compressed format. 
The memory of each node is initialized to zero and updated after each graph event. 
Given a node addition event of $v_i$, $v_i$'s message $\mathbf{m}_i^{node}(t)$ at time $t$ is computed as the concatenation of $i$'s raw features and memory: 
\begin{align}
    \mathbf{m}_i^{node}(t) &=\operatorname{MLP} \left(\left[\mathbf{h}_i(t^{\prime}) || \mathbf{x}_i(t) || \phi(t)\right]\right), 
\end{align}

\noindent where $\mathbf{h}_i\left(t^{\prime}\right)$ is $v_i$'s memory from time $t'$, \ie, the time of the previous interaction involving $v_i$. 
In the same manner, we obtain each community $s_j$'s message $\mathbf{m}_j(t)$ at $t$ given $s_j$'s event.

For an edge addition event involving $v_i$ and $s_j$, the edge's message $\mathbf{m}_i^{edge}(t)$ with respect to $v_i$ at $t$ is computed as: 
\begin{align}
\mathbf{m}_{i}^{edge}(t) &= \operatorname{MLP} \left(\left[\mathbf{h}_i(t^{\prime}) || \mathbf{h}_j(t^{\prime}) || \mathbf{x}_i(t) || \mathbf{x}_j(t) || \phi(t) \right] \right).
\end{align}
Similarly, we can obtain the edge's message $\mathbf{m}_j^{edge}(t)$ with respect to $s_j$ at $t$.

During batch training, multiple events in the same batch can be associated with the same nodes.
Therefore, we aggregate multiple messages of video $v_i$ and community $s_j$ from $t_1$ to $t_B$ through mean pooling, thus obtaining $\overline{\mathbf{m}}_i(t)$ and $\overline{\mathbf{m}}_j(t)$ as in ~\cite{rossi2020temporal}. 

Based on these messages, the memory embeddings of $v_i$ and $s_j$ are updated upon each event involving $v_i$ and $s_j$, respectively:
\begin{align}
    \mathbf{h}_i(t)=\operatorname{GRU}\left(\overline{\mathbf{m}}_i(t), \mathbf{h}_i(t^{\prime})\right),\\
    \mathbf{h}_j(t)=\operatorname{GRU}\left(\overline{\mathbf{m}}_j(t), \mathbf{h}_j(t^{\prime})  \right).
\end{align}

During prediction, we pass the representation $\mathbf{h}_i(t), \mathbf{h}_j(t)$ through multiple GNN layers to aggregate the features of each node from its neighbors on $G^D$
\begin{equation}
    \mathbf{\tilde{v}}_{i}^{t} = f_{\theta}(\mathbf{h}_i(t), \mathcal{G}^D), \quad
    \mathbf{\tilde{s}}_{j}^{t} = f_{\theta}(\mathbf{h}_j(t), \mathcal{G}^D),
    \label{eq:gnn}
\end{equation}
where $\mathbf{\tilde{v}}_{i}^{t}, \mathbf{\tilde{s}}_{j}^{t}$ are the transformed representation of $v_i, s_j$. The aggregation function $f_{\theta}(\cdot, \cdot)$ can be chosen from a wide range of GNN operators, such as GCN~\cite{kipf2016semi}, GraphSAGE~\cite{hamilton2017inductive}, TransformerConv~\cite{shimasked}, and GIN~\cite{xu2018powerful}. In practice, we employ a 2-layer Graph Attention Network (GAT)~\cite{velivckovic2018graph}.

\subsection{Training}
\label{sec:prediction}

We employ element-wise multiplication to calculate the score between each video $v_i$ and each community $s_j$ at time $t$:
\begin{align}
\hat{y}_{ij}^{t} &= \operatorname{MLP}(\mathbf{\tilde{v}}_{i}^{t} \odot \operatorname{MLP}(\mathbf{\tilde{s}}_{j}^{t})),
\label{eq:file_level_prediction}
\end{align}
where $\hat{y}_{ij}^{t}$ is the predicted score between $v_i$ and $s_j$. 
We train our model using the Bayesian Personalized Ranking (BPR)~\cite{rendle2009bpr} loss, which encourages the prediction of an observed interaction to be greater than an unobserved one:
\begin{align}
\mathcal{L}_{\text{BPR}} &=\sum_{(i, j^{+}, j^{-}, t)}-\ln (\operatorname{sigmoid}(\hat{y}_{ij^{+}}^{t} -\hat{y}_{ij^{-}}^{t})),
\label{eq:BPRLoss}
\end{align}
where $(i, j^{+}, j^{-}, t)$ denotes an example in the pairwise training data. $j^{+}$ indicates that one sharing of $v_i$ is observed in community $s_{j^{+}}$, and $j^{-}$ indicates an unobserved one. 

Furthermore, for the training of the community influence graph, we use the next item prediction objective. Given each $\mathcal{G}_i^{S}$, the loss function $\mathcal{L}_{\text{CE}}^{i}$ is defined as the cross-entropy of the predicted and ground-truth community that will propagate $v_i$ at the next timestamp:
\begin{equation}
\mathcal{L}_{\text{CE}}^{i}= \operatorname{CrossEntropy}(\mathbb{P}(s_{N+1} | v_i, P_i), \mathbf{y}_{N+1}),
\label{eq:CELoss}
\end{equation}
where $\mathbf{y}_{N+1} \in \mathbb{R}^{|\mathcal{S}|}$ is a one-hot vector that encodes the ground-truth community interacted at the next timestamp. 

The overall optimization objective is defined as follows: 
\begin{equation}
    \mathcal{L} = \mathcal{L}_{\text{BPR}} + \lambda_1 \sum_{i \in \mathcal{V}} \mathcal{L}_{\text{CE}}^{i} + \lambda_2 \|\Theta\|_2,
\end{equation}
where $\Theta$ denotes all trainable model parameters. $\lambda_1$ and $\lambda_2$ are hyperparameters in \model.

\section{Evaluation}

In this section, we conduct experiments to answer the following evaluation questions (EQs):

\begin{itemize}[leftmargin=*]
    \item {\bf (EQ1)} Does \model\ outperform the baseline models for the task of community-level information pathway prediction (Section~\ref{sec:warmstart})?
    \item {\bf (EQ2)} Does \model\ provide excellent inductive reasoning for cold-start videos (Section~\ref{sec:coldstart})?
    \item {\bf (EQ3)} What is the contribution of each component in \model\ (Section~\ref{sec:ablation})? 
    \item {\bf (EQ4)} Do community influence graphs (CIGs) constructed by \model\ manifest macroscopic influences (Section~\ref{sec:visualization})?
\end{itemize}

\begin{table*}
\centering
\caption{Performances of \model and 7 competitors for warm-start videos on the \textsf{Large} dataset. Values in bold and underline represent the best and the second best performances in each row, respectively. ``Impr.'' denotes the performance improvement of \model compared to the best baseline. 
}\label{tab:warm}
\vspace{-0.5cm}
\textbf{(a) Popular Subreddits}
\setlength{\tabcolsep}{4.2pt}
\begin{tabular}{c|cccccccc|c}
\toprule
& MF & NGCF & LightGCN & SVD-GCN & TiSASRec & TGAT & TGN & \model & Impr. \\
\midrule
NDCG@5 & 52.26 $\pm$ 0.52 & 52.97 $\pm$ 0.40 & 54.74 $\pm$ 0.55 & 56.85 $\pm$ 0.32 & 57.02 $\pm$ 0.44 & 56.85 $\pm$ 0.41 & \ul{57.28} $\pm$ 0.57 & \textbf{60.19} $\pm$ 0.38 & 5.1\% \\
Rec@5 & 71.98 $\pm$ 0.33 & 73.14 $\pm$ 0.25 & 73.56 $\pm$ 0.45 & 75.78 $\pm$ 0.40 & 76.00 $\pm$ 0.51 & 76.02 $\pm$ 0.58 & \ul{76.08} $\pm$ 0.43 & \textbf{78.32} $\pm$ 0.47 & 3.0\% \\
NDCG@10 & 60.15 $\pm$ 0.44 & 55.99 $\pm$ 0.54 & 56.81 $\pm$ 0.58 & 60.22 $\pm$ 0.49 & 60.34 $\pm$ 0.33 & 61.35 $\pm$ 0.31 & \ul{61.46} $\pm$ 0.40 & \textbf{63.89} $\pm$ 0.39 & 4.0\% \\
Rec@10 & 84.76 $\pm$ 0.23 & 84.82 $\pm$ 0.26 & 85.12 $\pm$ 0.49 & 85.19 $\pm$ 0.43 & 85.39 $\pm$ 0.38 & 85.34 $\pm$ 0.39 & \ul{85.56} $\pm$ 0.35 & \textbf{87.98} $\pm$ 0.50 & 2.8\% \\
MRR & 47.38 $\pm$ 0.51 & 52.52 $\pm$ 0.36 & 52.20 $\pm$ 0.35 & 53.85 $\pm$ 0.29 & 53.60 $\pm$ 0.46 & 53.58 $\pm$ 0.47 & \ul{55.82} $\pm$ 0.53 & \textbf{58.27} $\pm$ 0.36 & 4.4\% \\
\bottomrule
\end{tabular}
\textbf{(b) Non-popular Subreddits}
\setlength{\tabcolsep}{4.2pt}
\begin{tabular}{c|cccccccc|c}
\toprule
NDCG@5 & 13.54 $\pm$ 0.31 & 14.05 $\pm$ 0.48 & 15.42 $\pm$ 0.44 & 16.15 $\pm$ 0.55 & 16.72 $\pm$ 0.35 & 16.82 $\pm$ 0.43 & \ul{16.89} $\pm$ 0.52 & \textbf{18.04} $\pm$ 0.56 & 6.8\% \\
Rec@5 & 22.08 $\pm$ 0.38 & 22.19 $\pm$ 0.51 & 24.32 $\pm$ 0.38 & 25.44 $\pm$ 0.53 & 25.90 $\pm$ 0.36 & 25.95 $\pm$ 0.45 & \ul{25.99} $\pm$ 0.44 & \textbf{27.46} $\pm$ 0.40 & 5.7\% \\
NDCG@10 & 18.28 $\pm$ 0.59 & 18.50 $\pm$ 0.45 & 19.93 $\pm$ 0.50 & 20.70 $\pm$ 0.42 & 20.82 $\pm$ 0.39 & 21.26 $\pm$ 0.46 & \ul{21.42} $\pm$ 0.57 & \textbf{22.67} $\pm$ 0.37 & 5.8\% \\
Rec@10 & 36.03 $\pm$ 0.27 & 36.88 $\pm$ 0.38 & 38.39 $\pm$ 0.55 & 39.68 $\pm$ 0.46 & 39.62 $\pm$ 0.30 & 39.75 $\pm$ 0.31 & \ul{39.76} $\pm$ 0.38 & \textbf{41.88} $\pm$ 0.52 & 5.3\% \\
MRR & 15.17 $\pm$ 0.55 & 15.87 $\pm$ 0.57 & 16.96 $\pm$ 0.51 & 17.45 $\pm$ 0.58 & 17.75 $\pm$ 0.48 & 17.79 $\pm$ 0.49 & \ul{18.22} $\pm$ 0.41 & \textbf{19.27} $\pm$ 0.55 & 5.8\% \\
\bottomrule
\end{tabular}
\end{table*}
\vspace{-0.5cm}

\begin{table*}
\centering
\vspace{-0.3cm}
\caption{Performances of \model\ and 7 competitors for cold-start videos on the \textsf{Large} dataset. }
\label{tab:cold}
\vspace{-0.3cm}
\setlength{\tabcolsep}{4pt}
\textbf{(a) Popular Subreddits}
\begin{tabular}{c|cccccccc|c}
    \toprule
    & MF & NGCF & LightGCN & SVD-GCN & TiSASRec & TGAT & TGN & \model & \textbf{Impr.} \\
    \midrule
    NDCG@5 & 52.88 $\pm$ 0.43 & 56.28 $\pm$ 0.33 & 57.73 $\pm$ 0.45 & 58.07 $\pm$ 0.37 & 58.52 $\pm$ 0.39 & \ul{58.95} $\pm$ 0.53 & 58.79 $\pm$ 0.38 & \textbf{61.85} $\pm$ 0.38 & 4.9\% \\
    Rec@5 & 73.55 $\pm$ 0.42 & 74.83 $\pm$ 0.34 & 75.41 $\pm$ 0.28 & 76.31 $\pm$ 0.46 & 76.02 $\pm$ 0.55 & \ul{76.38} $\pm$ 0.57 & 76.35 $\pm$ 0.47 & \textbf{78.54} $\pm$ 0.53 & 2.8\% \\
    NDCG@10 & 56.73 $\pm$ 0.38 & 58.51 $\pm$ 0.52 & 60.13 $\pm$ 0.54 & 60.32 $\pm$ 0.37 & 60.22 $\pm$ 0.51 & \ul{61.03} $\pm$ 0.57 & 60.92 $\pm$ 0.39 & \textbf{64.08} $\pm$ 0.44 & 5.0\% \\
    Rec@10 & 84.12 $\pm$ 0.56 & 83.95 $\pm$ 0.58 & 83.84 $\pm$ 0.34 & 84.06 $\pm$ 0.37 & 83.85 $\pm$ 0.59 & \ul{84.23} $\pm$ 0.53 & 84.12 $\pm$ 0.49 & \textbf{86.89} $\pm$ 0.52 & 3.2\% \\
    MRR & 48.25 $\pm$ 0.45 & 51.17 $\pm$ 0.50 & 52.57 $\pm$ 0.47 & 53.33 $\pm$ 0.35 & 53.28 $\pm$ 0.32 & \ul{54.88} $\pm$ 0.49 & 54.64 $\pm$ 0.54 & \textbf{57.58} $\pm$ 0.36 & 4.9\% \\
    \bottomrule
\end{tabular}
\setlength{\tabcolsep}{4pt}
\textbf{(b) Non-Popular Subreddits}
\begin{tabular}{c|cccccccc|c}
    \toprule
    NDCG@5 & 10.64 $\pm$ 0.36 & 13.79 $\pm$ 0.39 & 14.34 $\pm$ 0.37 & 14.83 $\pm$ 0.33 & 15.13 $\pm$ 0.37 & 15.84 $\pm$ 0.52 & \ul{16.23} $\pm$ 0.32 & \textbf{17.61} $\pm$ 0.58 & 8.5\% \\
    Rec@5 & 15.92 $\pm$ 0.54 & 22.91 $\pm$ 0.57 & 25.24 $\pm$ 0.38 & 25.31 $\pm$ 0.37 & 25.38 $\pm$ 0.30 & \ul{25.49} $\pm$ 0.45 & 25.46 $\pm$ 0.59 & \textbf{27.06} $\pm$ 0.40 & 6.2\% \\
    NDCG@10 & 13.98 $\pm$ 0.51 & 18.33 $\pm$ 0.38 & 19.51 $\pm$ 0.52 & 19.44 $\pm$ 0.55 & 19.89 $\pm$ 0.50 & 20.66 $\pm$ 0.44 & \ul{20.79} $\pm$ 0.37 & \textbf{22.04} $\pm$ 0.45 & 6.0\% \\
    Rec@10 & 25.88 $\pm$ 0.41 & 37.30 $\pm$ 0.49 & 39.33 $\pm$ 0.41 & 39.70 $\pm$ 0.36 & 39.72 $\pm$ 0.54 & 40.06 $\pm$ 0.39 & \ul{40.17} $\pm$ 0.57 & \textbf{42.37} $\pm$ 0.32 & 5.5\% \\
    MRR & 12.42 $\pm$ 0.55 & 15.06 $\pm$ 0.38 & 15.83 $\pm$ 0.55 & 16.94 $\pm$ 0.59 & 17.03 $\pm$ 0.38 & \ul{17.58} $\pm$ 0.58 & 17.30 $\pm$ 0.52 & 18.64 $\pm$ 0.48 & 6.0\% \\
    \bottomrule
\end{tabular}
\end{table*}

\subsection{Experimental Setup}
\label{sec:experimentalSetup}

\subsubsection{Datasets}\label{sec:data-stat}
We construct 2 multi-modal datasets that provide the diffusion of YouTube videos on Reddit. Details can be found in Section~\ref{sec:data}. Table~\ref{tab:data} provides an overview of their statistics. 
To partition the datasets into train/validation/test sets, we used a 70/15/15 ratio based on the timestamps in a sequential manner. 
To ensure validity, we constructed the community influence graphs (CIGs) exclusively using the interactions from the training set to prevent any potential information leakage. 

\subsubsection{Baselines}

To evaluate the effectiveness of \model, we compare \model\ with seven baselines. 
We categorize them into four folds: (1) \emph{Matrix Factorization}, including MF~\cite{rendle2009bpr}; (2) \emph{Graph-based Recommendation}, including NGCF~\cite{wang2019neural}, LightGCN~\cite{he2020lightgcn}, and SVD-GCN~\cite{peng2022svd}; (3) \emph{Sequential Recommendation}, including TiSASRec~\cite{li2020time}; (4) \emph{Representation Learning on Temporal Graphs}, including TGAT~\cite{xu2020inductive}, and TGN~\cite{rossi2020temporal}.

\subsubsection{Metrics}

We measure the models' performances using three widely adopted metrics in the field of ranking systems: (1) recall@$K$, which measures the proportion of relevant items (\ie, ground truth) that are retrieved among the top-$K$ items; 
(2) normalized discounted cumulative gain (NDCG@$K$), which evaluates the ranking quality of the top-$K$ items, with a score of 1 assigned to the ideal ranking; 
(3) mean reciprocal rank (MRR), which computes the average reciprocal rank of the top-ranked relevant item. 
In this paper, we set $K$ to 5 and 10. Our evaluation procedure follows the established method~\cite{li2020time, he2017neural, elkahky2015multi} by randomly selecting 100 communities with no observed propagations of the video and ranking the test item among the 100 items. Additionally, we exclude any existing interactions in the training set from the test set. 


\subsubsection{Implementation Details}

We implemented \model\ in PyTorch \cite{paszke2019pytorch} and PyG~\cite{Fey/Lenssen/2019}. 
For a fair comparison, we set the embedding size to 64 in all methods including \model\ and perform Xavier initialization~\cite{glorot2010understanding} on the model parameters. 
We use Adam optimizer~\cite{kingma2015adam} with a batch size of 256. 
For the baseline models, the hyperparameters are set to the optimal values as reported in the original paper. 
For all models, we search the learning rate within the range of $[1e-4, 3e-4, 1e-3, 3e-3, 1e-2]$ and select the best setting. We set $\alpha=0.1$, $c=3$, $\lambda_1=1$, and $\lambda_2=1e-3$, respectively. $L$, the number of GNN layers in Community Influence Modeling (Section~\ref{sec:community}), is set to 4.


\subsection{Overall Performances}\label{sec:overall}

We conducted comparative experiments on 2 datasets to demonstrate the superiority of \model over the 7 baselines.
To this end, we grouped the videos into warm-start and cold-start videos.
We define warm-start and cold-start videos as videos with $\ge 2$ postings and 1 postings in the training phase, respectively. 
Furthermore, the number of videos posted in communities creates an imbalanced distribution. 
For instance, in the small dataset, more than 20\% of videos were posted on the \textit{two} most popular subreddits. 
Since it is relatively trivial to make predictions for such popular subreddits, we split subreddits into popular (\ie, top 25 percentile subreddits where YouTube videos are posted most frequently) and non-popular (\ie, the rest of the subreddits). 
The results are partitioned with respect to whether the target community is a popular subreddit or a non-popular subreddit. 

\subsubsection{Warm-Start Prediction}\label{sec:warmstart}
Tables~\ref{tab:warm}(a)-(b) show the results for warm-start prediction for popular and non-popular subreddits, respectively, on the \textsf{Large} dataset.
The results for the \textsf{Small} dataset can be found in Appendix~\ref{app:evaluation}.
We observe that \model\ consistently and significantly outperforms all baselines on both datasets for both groups of subreddits. On the \textsf{Large} dataset, \model\ outperforms the best baseline by 5.1\% on NDCG@5 and 4.4\% on MRR for the popular communities, as well as 6.8\% on NDCG@5 and 5.8\% on MRR for non-popular communities, respectively. 
On the \textsf{Small} dataset, \model\ outperforms the best competitor by 8.6\% and 7.5\% on the two metrics for popular communities, and 12.9\% and 18.8\% for popular communities, respectively. 
Our results demonstrate the effectiveness of \model\ in the task of \problem.
Moreover, we observe that representation learning methods on temporal graphs (\ie, TGAT and TGN) outperform all other baselines. This observation underscores the importance of considering temporal information in predicting information pathways. 

\subsubsection{Cold-Start Prediction}\label{sec:coldstart}

As the content sharing network evolves, the emergence and spread of new content to a diverse range of communities presents considerable challenges for \problem, particularly in cold-start scenarios where historical propagation of videos is absent. 
Thus, the prediction problem becomes: \emph{given a video that has only 1 propagation, how can we predict its second propagation?} 
Tables~\ref{tab:cold}(a)-(b) show the performances of seven baselines and \model for popular and non-popular subreddits, respectively, on the \textsf{Large} dataset.
The results for the \textsf{Small} dataset can be found in Appendix~\ref{app:evaluation}.
We observe that \model\ is able to achieve even greater performance improvements in the cold-start scenario through its inductive reasoning capability, consistently outperforming all competitors on both datasets for both groups of subreddits. Moreover, from Table~\ref{tab:cold}(a), we observed that when the cold-start videos are propagated to popular communities, predicting these flows is relatively straightforward for all the models, including \model.
On the other hand, the results in Table~\ref{tab:cold}(b) show that predicting the flow of cold-start videos to less popular communities is a more challenging task. Despite this, \model\ still shows the best performance. These results encourage further investigation into such flows, which we consider to be a potential area of future work.

\subsection{Ablation Studies}\label{sec:ablation}
\begin{figure}
    \centering
    \includegraphics[width=0.42\textwidth]{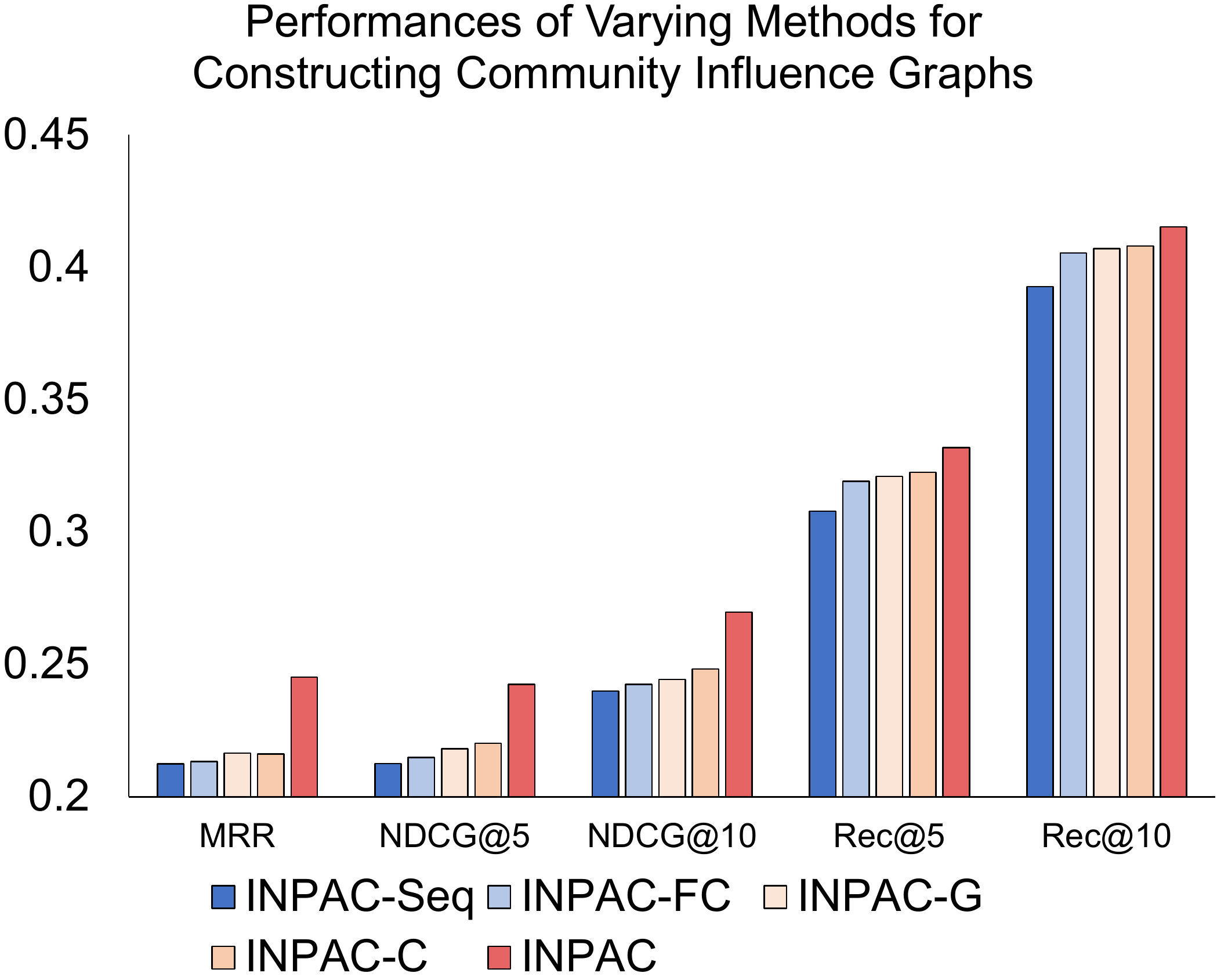}
    \caption{Performances of different methods for constructing the community influence graph (CIG) on the \textsf{Small} dataset. 
    }
    \vspace{-0.3cm}
    \label{fig:CommunityInfluenceGraphSmall}
    \vspace{-0.3cm}
\end{figure}

We validate the effectiveness of the design choices in \model.
In Section~\ref{sec:community}, we designed a way to construct community influence graphs (CIGs) by considering the time that videos were propagated in communities.
To evaluate this design, we made 4 variants of \model:
\model-Seq connects the community nodes sequentially, \ie, we create a directed edge from $s_j$ to $s_k$ if they are adjacent in the corresponding propagation sequence $P_i$. 
\model-FC establishes connections in a fully-connected manner, meaning that an edge is created between $s_j$ and $s_k$ if $s_j$ precedes $s_k$ in $P_i$. 
\model-G adopts the graph construction method of CIG as suggested by the GAINRec model~\cite{meng2023recognize}. 
\model-C omits any content information about the video and its channel. Specifically, the video embeddings $\mathbf{v}_{i}$ and the channel embedding $\mathbf{c}_{\rho(i)}$ in Eq.~(\ref{eq:AggregateEmbeddings}) are randomly initialized. 

From Figure~\ref{fig:CommunityInfluenceGraphSmall}, we observe that 
\model-Seq exhibits the lowest performances. This result can be attributed to the limitations of the sequential connection method, which fails to capture the underlying influencing relationships between communities as manifested by the sharing events. On the other hand, 
\model-FC performs better than \model-Seq in terms of Rec@5 and Rec@10. However, the fully-connected method can potentially lead to spurious correlations. 
Perhaps surprisingly, \model-C outperforms both \model-Seq, \model-FC, and \model-G on most metrics, suggesting that the model can still achieve remarkable performance in the absence of content features, given that the Community Influence Graphs (CIGs) are properly constructed and modeled.  This has broader implications for its applicability to other types of information with less available content, such as short online posts or URLs to misinformative websites. The superior performance of \model-C highlights the importance of the construction of the Community Influence Graph (CIG) in our approach. The CIG captures the interactions and influence patterns among communities, which is a crucial aspect when modeling the spread of information in online social networks. By focusing on the underlying social structures, our method is able to identify and predict the propagation of information more effectively than solely relying on content features. 
Overall, the method employed by \model achieves the best performances, demonstrating the effectiveness of our graph construction approach.



\subsection{Analysis of CIG}\label{sec:visualization}

\begin{figure}[t]
    \centering
    \includegraphics[width=0.47\textwidth]{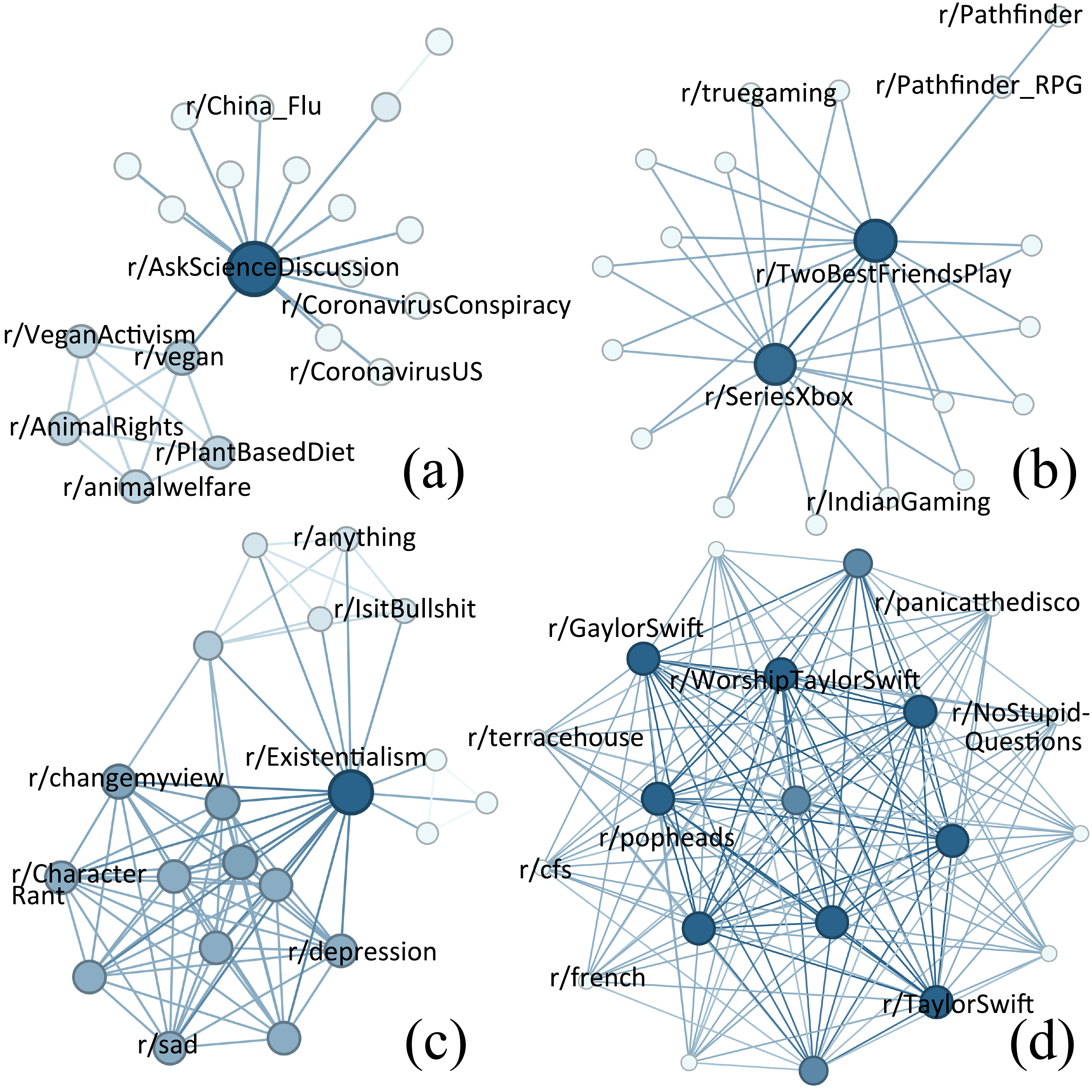}
    \caption{Community Influence Graphs (CIGs) of 4 different videos, all of which were propagated in exactly 20 communities. (a) How Wildlife Trade is Linked to Coronavirus; (b) Black Myth: Wukong - Official 13 Minutes Gameplay Trailer; (c) Thought experiment ``BRAIN IN A VAT''; (d) Taylor Swift - ME! Node sizes and colors indicate the node degrees. Edge colors indicate the edge weights. 
    }
    \label{fig:visualization}
    \vspace{-0.5cm}
\end{figure}

Figure~\ref{fig:visualization} presents the visualization of Community Influence Graphs (CIGs) for 4 videos with different topics (Section~\ref{sec:community}). Each video was propagated in exactly 20 communities. The node colors and sizes in the graphs represent the node degrees, while edge colors indicate the edge weights. 
We observe that CIGs generated from different videos demonstrate diverse connectivities and structures. 
We categorized the CIGs into two groups: (1) CIGs with multiple clusters, such as Figures~\ref{fig:visualization}(a)(c); and (2) CIGs with a single cluster, such as Figures~\ref{fig:visualization}(b)(d).

Regarding the CIGs with multiple clusters, we analyzed the differences between the clusters and the factors that contributed to the video spreading across different clusters. In Figures~\ref{fig:visualization}(a)(c), the videos were first posted in highly active communities. As the videos gained visibility over time, they spread to different clusters of communities. For instance, in Figure~\ref{fig:visualization}(a), the video was initially shared in \textsf{r/AskScienceDiscussion}, a community focused on in-depth scientific discussions, which aligned with the video's original purpose. Subsequently, as the video gained popularity, it was shared by distinct users in highly active COVID-19 related communities such as \textsf{r/CoronavirusUS} and \textsf{r/China\_Flu}. Furthermore, the video also sheds light on the poor living conditions of animals in produce markets, where animals are confined in stacked cages and subjected to unsanitary conditions, evoking sympathy among viewers regarding animal welfare. As a result, the video was shared in 5 topically similar communities related to vegetarianism and animal welfare, including \textsf{r/Vegan}, \textsf{r/VeganActivism}, \textsf{r/PlantBasedDiet}, \textsf{r/AnimalRights}, and \textsf{r/animalwelfare}. 
In fact, the same group of users spread the video to multiple semantically similar communities potentially due to overlapping interests. 
Our \model model successfully models these correlated sharing behaviors as a 5-clique. 

On the other hand, the CIGs in Figures~\ref{fig:visualization}(b)(d) exhibit a single cluster. We manually examined how these videos spread to communities with less obvious topical similarities. For instance, in Figure~\ref{fig:visualization}(d), the video first appeared in subreddits like \textsf{r/WorshipTaylorSwift}, a popular subreddits centered around the famous singer Taylor Swift, which directly relates to the posted video. Subsequently, the video propagated to multiple semantically distinct communities at different time periods. These communities included \textsf{r/terracehouse}, a subreddit about the reality TV show Terrace House, where users compared Taylor Swift's songs with the show's theme song and other famous singers' songs. Another example is \textsf{r/NoStupidQuestions}, a subreddit for discussing a wide range of curious questions, where a user shared this video and questioned people's obsession with Taylor Swift. Our key findings are as follows: 
\begin{itemize}[leftmargin=*]
    \item Initially, online content tends to be shared within communities that closely match its topic. As the content gains popularity, it gradually spreads to multiple communities with a broader range of topics.
    \item Content is shared within topically similar communities in a short period, regardless of whether it is shared by the same user or different users. This observation aligns with previous studies~\cite{shin2018diffusion} that found faster/slower information diffusion among topically similar/distant communities, respectively.
    \item There exist ``Super spreaders'' on online platforms who actively engage in and disseminate content across multiple topically diverse communities. For example, we identified a user who played a significant role in spreading the video in Figure~\ref{fig:visualization}(a) across vegetarian-related subreddits. This user has posted a total of 118 YouTube videos, with 67 shared in vegetarian-related communities. Another similar observation from Figure~\ref{fig:visualization}(c) is a user who actively contributed to communities about emotions, philosophy, Marvel Comics, and anime before eventually spreading the video among depression-related subreddits.
\end{itemize}


\section{Related Works}


\subsection{Information Diffusion}

Modeling the spread of information in online social networks has been a challenging task. Previous works have investigated information diffusion on social media~\cite{gomez2012inferring, el2022twhin}, prediction of popularity~\cite{cao2020popularity}, social influence~\cite{qiu2018deepinf, leung2019personalized}, and topological analysis of follower networks~\cite{kwak2010twitter, song2022friend} for information sharing. 
While these studies cover a broad spectrum of social interactions in online communities, they generally focus on user-level influence and interactions.  
Research has shown that the dissemination of information within a community is different from that at the individual level~\cite{chen2022ctl, zhou2020variational, myers2014bursty, pacheco2021uncovering}. In this sense, diffusion models have been used to understand the spread of ideas, information and influence on social and information networks~\cite{li2017deepcas, ng2022cross}.
Our study differs from the prior studies in its methodology as it endeavors to delve into the intricacies of community-level interactions.

\subsection{Graph Neural Networks}

Graph Neural Networks (GNNs)~\cite{kipf2016semi, zhu2021deep, cui2022positional, zhu2022structure, dong2022edits, dong2021individual, dong2023reliant} have received increased attention in recent years due to their exceptional capacity to model complex, non-Euclidean graph structures. 
Recently, GNNs have achieved state-of-the-art performances in various applications, including recommendation~\cite{hao2020p, wang2022multi, dwivedi2022affective, zhang2023survey,KimLSK22,KongKJ0LPK22, mao2022whole}, user modeling~\cite{huang2021coupled, zhang2023efficiently, guo2022evolutionary}, and social influence estimation~\cite{qiu2018deepinf, leung2019personalized, zhang2021mining}. 
These methods typically structure events into interaction graphs and leverage high-order relationships to derive node/edge attributes~\cite{jin2022code, hui2021trajnet, peng2022svd, jung2021learning,LeeL0K22}. 
Recently, dynamic graph models~\cite{wang2021inductive, kumar2019predicting, xu2020inductive, kazemi2020representation,YooLSK23} have emerged as powerful tools for various tasks, \eg, node classification, link prediction, and representation learning. 
The \problem problem can be modeled using dynamic networks~\cite{TGC_ML, S2T_ML, MNCI_ML_SIGIR} in which time-dependent representations of videos and communities are learned to infer future interactions. 

\section{Discussion and Conclusion}
Inference of community influence pathways can provide important information about the structure and dynamics of online platforms and the resulting information flow in the platform. 
This work created and utilized this influence graph in a dynamic graph framework \model to predict the flow of YouTube videos across Reddit communities (subreddits). 
Some shortcomings of this work include: (i) studying only YouTube-Reddit data and (ii) difficulty in the validation of the inferred influence graph.
Future work includes alternate approaches to generate and validate influence graphs, creation of new dynamic graph models to predict information flow, and using multi-platform data. 


\begin{acks}
This research/material is based upon work supported in part by NSF grants CNS-2154118, IIS-2027689, ITE-2137724, ITE-2230692, CNS-2239879, Defense Advanced Research Projects Agency (DARPA) under Agreement No. HR00112290102 (subcontract No. PO70745), and funding from Microsoft, Google, and Adobe Inc. Any opinions, findings, and conclusions or recommendations expressed in this material are those of the author(s) and do not necessarily reflect the position or policy of DARPA, DoD, SRI International, NSF and no official endorsement should be inferred. We thank the reviewers for their comments. 
\end{acks}

\bibliographystyle{ACM-Reference-Format}
\bibliography{sample-base}


\begin{thebibliography}{128}


\ifx \showCODEN    \undefined \def \showCODEN     #1{\unskip}     \fi
\ifx \showDOI      \undefined \def \showDOI       #1{#1}\fi
\ifx \showISBNx    \undefined \def \showISBNx     #1{\unskip}     \fi
\ifx \showISBNxiii \undefined \def \showISBNxiii  #1{\unskip}     \fi
\ifx \showISSN     \undefined \def \showISSN      #1{\unskip}     \fi
\ifx \showLCCN     \undefined \def \showLCCN      #1{\unskip}     \fi
\ifx \shownote     \undefined \def \shownote      #1{#1}          \fi
\ifx \showarticletitle \undefined \def \showarticletitle #1{#1}   \fi
\ifx \showURL      \undefined \def \showURL       {\relax}        \fi
\providecommand\bibfield[2]{#2}
\providecommand\bibinfo[2]{#2}
\providecommand\natexlab[1]{#1}
\providecommand\showeprint[2][]{arXiv:#2}

\bibitem[Ahmed et~al\mbox{.}(2020)]%
        {ahmed2020covid}
\bibfield{author}{\bibinfo{person}{Wasim Ahmed}, \bibinfo{person}{Josep
  Vidal-Alaball}, \bibinfo{person}{Joseph Downing},
  \bibinfo{person}{Francesc~L{\'o}pez Segu{\'\i}}, {et~al\mbox{.}}}
  \bibinfo{year}{2020}\natexlab{}.
\newblock \showarticletitle{COVID-19 and the 5G conspiracy theory: social
  network analysis of Twitter data}.
\newblock \bibinfo{journal}{\emph{Journal of Medical Internet Research}}
  \bibinfo{volume}{22}, \bibinfo{number}{5} (\bibinfo{year}{2020}),
  \bibinfo{pages}{e19458}.
\newblock


\bibitem[Brin and Page(1998)]%
        {brin1998anatomy}
\bibfield{author}{\bibinfo{person}{Sergey Brin} {and} \bibinfo{person}{Lawrence
  Page}.} \bibinfo{year}{1998}\natexlab{}.
\newblock \showarticletitle{The anatomy of a large-scale hypertextual web
  search engine}.
\newblock \bibinfo{journal}{\emph{Computer networks and ISDN systems}}
  \bibinfo{volume}{30}, \bibinfo{number}{1-7} (\bibinfo{year}{1998}),
  \bibinfo{pages}{107--117}.
\newblock


\bibitem[Cao et~al\mbox{.}(2020)]%
        {cao2020popularity}
\bibfield{author}{\bibinfo{person}{Qi Cao}, \bibinfo{person}{Huawei Shen},
  \bibinfo{person}{Jinhua Gao}, \bibinfo{person}{Bingzheng Wei}, {and}
  \bibinfo{person}{Xueqi Cheng}.} \bibinfo{year}{2020}\natexlab{}.
\newblock \showarticletitle{Popularity prediction on social platforms with
  coupled graph neural networks}. In \bibinfo{booktitle}{\emph{WSDM}}.
  \bibinfo{pages}{70--78}.
\newblock


\bibitem[Chang et~al\mbox{.}(2016)]%
        {chang2016positive}
\bibfield{author}{\bibinfo{person}{Shiyu Chang}, \bibinfo{person}{Yang Zhang},
  \bibinfo{person}{Jiliang Tang}, \bibinfo{person}{Dawei Yin},
  \bibinfo{person}{Yi Chang}, \bibinfo{person}{Mark~A Hasegawa-Johnson}, {and}
  \bibinfo{person}{Thomas~S Huang}.} \bibinfo{year}{2016}\natexlab{}.
\newblock \showarticletitle{Positive-Unlabeled Learning in Streaming
  Networks.}. In \bibinfo{booktitle}{\emph{KDD}}. \bibinfo{pages}{755--764}.
\newblock


\bibitem[Chen et~al\mbox{.}(2022b)]%
        {chen2022ctl}
\bibfield{author}{\bibinfo{person}{Jinyin Chen}, \bibinfo{person}{Xiaodong Xu},
  \bibinfo{person}{Lihong Chen}, \bibinfo{person}{Zhongyuan Ruan},
  \bibinfo{person}{Zhaoyan Ming}, {and} \bibinfo{person}{Yi Liu}.}
  \bibinfo{year}{2022}\natexlab{b}.
\newblock \showarticletitle{CTL-DIFF: Control Information Diffusion in Social
  Network by Structure Optimization}.
\newblock \bibinfo{journal}{\emph{IEEE Transactions on Computational Social
  Systems}} (\bibinfo{year}{2022}).
\newblock


\bibitem[Chen and Wong(2020)]%
        {chen2020handling}
\bibfield{author}{\bibinfo{person}{Tianwen Chen} {and} \bibinfo{person}{Raymond
  Chi-Wing Wong}.} \bibinfo{year}{2020}\natexlab{}.
\newblock \showarticletitle{Handling information loss of graph neural networks
  for session-based recommendation}. In \bibinfo{booktitle}{\emph{KDD}}.
  \bibinfo{pages}{1172--1180}.
\newblock


\bibitem[Chen et~al\mbox{.}(2019)]%
        {chen2019joint}
\bibfield{author}{\bibinfo{person}{Wanyu Chen}, \bibinfo{person}{Fei Cai},
  \bibinfo{person}{Honghui Chen}, {and} \bibinfo{person}{Maarten~De Rijke}.}
  \bibinfo{year}{2019}\natexlab{}.
\newblock \showarticletitle{Joint neural collaborative filtering for
  recommender systems}.
\newblock \bibinfo{journal}{\emph{TOIS}} \bibinfo{volume}{37},
  \bibinfo{number}{4} (\bibinfo{year}{2019}), \bibinfo{pages}{1--30}.
\newblock


\bibitem[Chen et~al\mbox{.}(2022a)]%
        {chen2022cross}
\bibfield{author}{\bibinfo{person}{Yixuan Chen}, \bibinfo{person}{Dongsheng
  Li}, \bibinfo{person}{Peng Zhang}, \bibinfo{person}{Jie Sui},
  \bibinfo{person}{Qin Lv}, \bibinfo{person}{Lu Tun}, {and} \bibinfo{person}{Li
  Shang}.} \bibinfo{year}{2022}\natexlab{a}.
\newblock \showarticletitle{Cross-modal ambiguity learning for multimodal fake
  news detection}. In \bibinfo{booktitle}{\emph{TheWebConf}}.
  \bibinfo{pages}{2897--2905}.
\newblock


\bibitem[Cui et~al\mbox{.}(2022)]%
        {cui2022positional}
\bibfield{author}{\bibinfo{person}{Hejie Cui}, \bibinfo{person}{Zijie Lu},
  \bibinfo{person}{Pan Li}, {and} \bibinfo{person}{Carl Yang}.}
  \bibinfo{year}{2022}\natexlab{}.
\newblock \showarticletitle{On positional and structural node features for
  graph neural networks on non-attributed graphs}. In
  \bibinfo{booktitle}{\emph{CIKM}}. \bibinfo{pages}{3898--3902}.
\newblock


\bibitem[Devlin et~al\mbox{.}(2019)]%
        {devlin2019bert}
\bibfield{author}{\bibinfo{person}{Jacob Devlin}, \bibinfo{person}{Ming-Wei
  Chang}, \bibinfo{person}{Kenton Lee}, {and} \bibinfo{person}{Kristina
  Toutanova}.} \bibinfo{year}{2019}\natexlab{}.
\newblock \showarticletitle{BERT: Pre-training of Deep Bidirectional
  Transformers for Language Understanding}. In
  \bibinfo{booktitle}{\emph{NAACL}}. \bibinfo{pages}{4171--4186}.
\newblock


\bibitem[Dong et~al\mbox{.}(2021)]%
        {dong2021individual}
\bibfield{author}{\bibinfo{person}{Yushun Dong}, \bibinfo{person}{Jian Kang},
  \bibinfo{person}{Hanghang Tong}, {and} \bibinfo{person}{Jundong Li}.}
  \bibinfo{year}{2021}\natexlab{}.
\newblock \showarticletitle{Individual fairness for graph neural networks: A
  ranking based approach}. In \bibinfo{booktitle}{\emph{KDD}}.
  \bibinfo{pages}{300--310}.
\newblock


\bibitem[Dong et~al\mbox{.}(2022)]%
        {dong2022edits}
\bibfield{author}{\bibinfo{person}{Yushun Dong}, \bibinfo{person}{Ninghao Liu},
  \bibinfo{person}{Brian Jalaian}, {and} \bibinfo{person}{Jundong Li}.}
  \bibinfo{year}{2022}\natexlab{}.
\newblock \showarticletitle{Edits: Modeling and mitigating data bias for graph
  neural networks}. In \bibinfo{booktitle}{\emph{TheWebConf}}.
  \bibinfo{pages}{1259--1269}.
\newblock


\bibitem[Dong et~al\mbox{.}(2023)]%
        {dong2023reliant}
\bibfield{author}{\bibinfo{person}{Yushun Dong}, \bibinfo{person}{Bilichi
  Zhang}, \bibinfo{person}{Yiling Yuan}, \bibinfo{person}{Na Zou},
  \bibinfo{person}{Qi Wang}, {and} \bibinfo{person}{Jundong Li}.}
  \bibinfo{year}{2023}\natexlab{}.
\newblock \showarticletitle{Reliant: Fair knowledge distillation for graph
  neural networks}. In \bibinfo{booktitle}{\emph{SDM}}.
  \bibinfo{pages}{154--162}.
\newblock


\bibitem[Dosovitskiy et~al\mbox{.}(2021)]%
        {dosovitskiyimage}
\bibfield{author}{\bibinfo{person}{Alexey Dosovitskiy}, \bibinfo{person}{Lucas
  Beyer}, \bibinfo{person}{Alexander Kolesnikov}, \bibinfo{person}{Dirk
  Weissenborn}, \bibinfo{person}{Xiaohua Zhai}, \bibinfo{person}{Thomas
  Unterthiner}, \bibinfo{person}{Mostafa Dehghani}, \bibinfo{person}{Matthias
  Minderer}, \bibinfo{person}{Georg Heigold}, {and} \bibinfo{person}{Sylvain
  Gelly}.} \bibinfo{year}{2021}\natexlab{}.
\newblock \showarticletitle{An Image is Worth 16x16 Words: Transformers for
  Image Recognition at Scale}. In \bibinfo{booktitle}{\emph{ICLR}}.
\newblock


\bibitem[Dwivedi-Yu et~al\mbox{.}(2022)]%
        {dwivedi2022affective}
\bibfield{author}{\bibinfo{person}{Jane Dwivedi-Yu}, \bibinfo{person}{Yi-Chia
  Wang}, \bibinfo{person}{Lijing Qin}, \bibinfo{person}{Cristian
  Canton-Ferrer}, {and} \bibinfo{person}{Alon~Y Halevy}.}
  \bibinfo{year}{2022}\natexlab{}.
\newblock \showarticletitle{Affective Signals in a Social Media Recommender
  System}. In \bibinfo{booktitle}{\emph{KDD}}. \bibinfo{pages}{2831--2841}.
\newblock


\bibitem[El-Kishky et~al\mbox{.}(2022)]%
        {el2022twhin}
\bibfield{author}{\bibinfo{person}{Ahmed El-Kishky}, \bibinfo{person}{Thomas
  Markovich}, \bibinfo{person}{Serim Park}, \bibinfo{person}{Chetan Verma},
  \bibinfo{person}{Baekjin Kim}, \bibinfo{person}{Ramy Eskander},
  \bibinfo{person}{Yury Malkov}, \bibinfo{person}{Frank Portman},
  \bibinfo{person}{Sof{\'\i}a Samaniego}, \bibinfo{person}{Ying Xiao},
  {et~al\mbox{.}}} \bibinfo{year}{2022}\natexlab{}.
\newblock \showarticletitle{Twhin: Embedding the twitter heterogeneous
  information network for personalized recommendation}. In
  \bibinfo{booktitle}{\emph{KDD}}. \bibinfo{pages}{2842--2850}.
\newblock


\bibitem[Elkahky et~al\mbox{.}(2015)]%
        {elkahky2015multi}
\bibfield{author}{\bibinfo{person}{Ali~Mamdouh Elkahky}, \bibinfo{person}{Yang
  Song}, {and} \bibinfo{person}{Xiaodong He}.} \bibinfo{year}{2015}\natexlab{}.
\newblock \showarticletitle{A multi-view deep learning approach for cross
  domain user modeling in recommendation systems}. In
  \bibinfo{booktitle}{\emph{TheWebConf}}. \bibinfo{pages}{278--288}.
\newblock


\bibitem[Fey and Lenssen(2019)]%
        {Fey/Lenssen/2019}
\bibfield{author}{\bibinfo{person}{Matthias Fey} {and} \bibinfo{person}{Jan~E.
  Lenssen}.} \bibinfo{year}{2019}\natexlab{}.
\newblock \showarticletitle{Fast Graph Representation Learning with {PyTorch
  Geometric}}. In \bibinfo{booktitle}{\emph{ICLR Workshop on Representation
  Learning on Graphs and Manifolds}}.
\newblock


\bibitem[Fiesler et~al\mbox{.}(2018)]%
        {fiesler2018reddit}
\bibfield{author}{\bibinfo{person}{Casey Fiesler}, \bibinfo{person}{Joshua
  McCann}, \bibinfo{person}{Kyle Frye}, \bibinfo{person}{Jed~R Brubaker},
  {et~al\mbox{.}}} \bibinfo{year}{2018}\natexlab{}.
\newblock \showarticletitle{Reddit rules! characterizing an ecosystem of
  governance}. In \bibinfo{booktitle}{\emph{ICWSM}}.
\newblock


\bibitem[Firdaus et~al\mbox{.}(2018)]%
        {firdaus2018retweet}
\bibfield{author}{\bibinfo{person}{Syeda~Nadia Firdaus}, \bibinfo{person}{Chen
  Ding}, {and} \bibinfo{person}{Alireza Sadeghian}.}
  \bibinfo{year}{2018}\natexlab{}.
\newblock \showarticletitle{Retweet: A popular information diffusion
  mechanism--A survey paper}.
\newblock \bibinfo{journal}{\emph{Online Social Networks and Media}}
  \bibinfo{volume}{6} (\bibinfo{year}{2018}), \bibinfo{pages}{26--40}.
\newblock


\bibitem[Gasteiger et~al\mbox{.}(2018)]%
        {gasteiger2018predict}
\bibfield{author}{\bibinfo{person}{Johannes Gasteiger},
  \bibinfo{person}{Aleksandar Bojchevski}, {and} \bibinfo{person}{Stephan
  G{\"u}nnemann}.} \bibinfo{year}{2018}\natexlab{}.
\newblock \showarticletitle{Predict then Propagate: Graph Neural Networks meet
  Personalized PageRank}. In \bibinfo{booktitle}{\emph{ICLR}}.
\newblock


\bibitem[Glorot and Bengio(2010)]%
        {glorot2010understanding}
\bibfield{author}{\bibinfo{person}{Xavier Glorot} {and} \bibinfo{person}{Yoshua
  Bengio}.} \bibinfo{year}{2010}\natexlab{}.
\newblock \showarticletitle{Understanding the difficulty of training deep
  feedforward neural networks}. In \bibinfo{booktitle}{\emph{AISTATS}}.
  \bibinfo{pages}{249--256}.
\newblock


\bibitem[Gomez-Rodriguez et~al\mbox{.}(2012)]%
        {gomez2012inferring}
\bibfield{author}{\bibinfo{person}{Manuel Gomez-Rodriguez},
  \bibinfo{person}{Jure Leskovec}, {and} \bibinfo{person}{Andreas Krause}.}
  \bibinfo{year}{2012}\natexlab{}.
\newblock \showarticletitle{Inferring networks of diffusion and influence}.
\newblock \bibinfo{journal}{\emph{TKDD}} \bibinfo{volume}{5},
  \bibinfo{number}{4} (\bibinfo{year}{2012}), \bibinfo{pages}{1--37}.
\newblock


\bibitem[Graves and Graves(2012)]%
        {graves2012long}
\bibfield{author}{\bibinfo{person}{Alex Graves} {and} \bibinfo{person}{Alex
  Graves}.} \bibinfo{year}{2012}\natexlab{}.
\newblock \showarticletitle{Long short-term memory}.
\newblock \bibinfo{journal}{\emph{Supervised sequence labelling with recurrent
  neural networks}} (\bibinfo{year}{2012}), \bibinfo{pages}{37--45}.
\newblock


\bibitem[Gulati et~al\mbox{.}(2020)]%
        {gulati2020conformer}
\bibfield{author}{\bibinfo{person}{Anmol Gulati}, \bibinfo{person}{James Qin},
  \bibinfo{person}{Chung-Cheng Chiu}, \bibinfo{person}{Niki Parmar},
  \bibinfo{person}{Yu Zhang}, \bibinfo{person}{Jiahui Yu}, \bibinfo{person}{Wei
  Han}, \bibinfo{person}{Shibo Wang}, \bibinfo{person}{Zhengdong Zhang},
  \bibinfo{person}{Yonghui Wu}, {et~al\mbox{.}}}
  \bibinfo{year}{2020}\natexlab{}.
\newblock \showarticletitle{Conformer: Convolution-augmented Transformer for
  Speech Recognition}.
\newblock \bibinfo{journal}{\emph{Interspeech}} (\bibinfo{year}{2020}),
  \bibinfo{pages}{5036--5040}.
\newblock


\bibitem[Guo et~al\mbox{.}(2022)]%
        {guo2022evolutionary}
\bibfield{author}{\bibinfo{person}{Jiayan Guo}, \bibinfo{person}{Peiyan Zhang},
  \bibinfo{person}{Chaozhuo Li}, \bibinfo{person}{Xing Xie},
  \bibinfo{person}{Yan Zhang}, {and} \bibinfo{person}{Sunghun Kim}.}
  \bibinfo{year}{2022}\natexlab{}.
\newblock \showarticletitle{Evolutionary Preference Learning via Graph Nested
  GRU ODE for Session-based Recommendation}. In
  \bibinfo{booktitle}{\emph{CIKM}}. \bibinfo{pages}{624--634}.
\newblock


\bibitem[Guo et~al\mbox{.}(2019)]%
        {guo2019streaming}
\bibfield{author}{\bibinfo{person}{Lei Guo}, \bibinfo{person}{Hongzhi Yin},
  \bibinfo{person}{Qinyong Wang}, \bibinfo{person}{Tong Chen},
  \bibinfo{person}{Alexander Zhou}, {and} \bibinfo{person}{Nguyen Quoc
  Viet~Hung}.} \bibinfo{year}{2019}\natexlab{}.
\newblock \showarticletitle{Streaming session-based recommendation}. In
  \bibinfo{booktitle}{\emph{KDD}}. \bibinfo{pages}{1569--1577}.
\newblock


\bibitem[Gustafsson(2010)]%
        {gustafsson2010time}
\bibfield{author}{\bibinfo{person}{Nils Gustafsson}.}
  \bibinfo{year}{2010}\natexlab{}.
\newblock \showarticletitle{This time it’s personal: Social networks, viral
  politics and identity management}.
\newblock In \bibinfo{booktitle}{\emph{Emerging practices in cyberculture and
  social networking}}. \bibinfo{publisher}{Brill}, \bibinfo{pages}{1--23}.
\newblock


\bibitem[Hamilton et~al\mbox{.}(2017)]%
        {hamilton2017inductive}
\bibfield{author}{\bibinfo{person}{Will Hamilton}, \bibinfo{person}{Zhitao
  Ying}, {and} \bibinfo{person}{Jure Leskovec}.}
  \bibinfo{year}{2017}\natexlab{}.
\newblock \showarticletitle{Inductive representation learning on large graphs}.
\newblock \bibinfo{journal}{\emph{NIPS}}  \bibinfo{volume}{30}
  (\bibinfo{year}{2017}).
\newblock


\bibitem[Hao et~al\mbox{.}(2020)]%
        {hao2020p}
\bibfield{author}{\bibinfo{person}{Junheng Hao}, \bibinfo{person}{Tong Zhao},
  \bibinfo{person}{Jin Li}, \bibinfo{person}{Xin~Luna Dong},
  \bibinfo{person}{Christos Faloutsos}, \bibinfo{person}{Yizhou Sun}, {and}
  \bibinfo{person}{Wei Wang}.} \bibinfo{year}{2020}\natexlab{}.
\newblock \showarticletitle{P-companion: A principled framework for diversified
  complementary product recommendation}. In \bibinfo{booktitle}{\emph{CIKM}}.
  \bibinfo{pages}{2517--2524}.
\newblock


\bibitem[Harper and Konstan(2015)]%
        {harper2015movielens}
\bibfield{author}{\bibinfo{person}{F~Maxwell Harper} {and}
  \bibinfo{person}{Joseph~A Konstan}.} \bibinfo{year}{2015}\natexlab{}.
\newblock \showarticletitle{The movielens datasets: History and context}.
\newblock \bibinfo{journal}{\emph{TIIS}} \bibinfo{volume}{5},
  \bibinfo{number}{4} (\bibinfo{year}{2015}), \bibinfo{pages}{1--19}.
\newblock


\bibitem[He et~al\mbox{.}(2016)]%
        {he2016deep}
\bibfield{author}{\bibinfo{person}{Kaiming He}, \bibinfo{person}{Xiangyu
  Zhang}, \bibinfo{person}{Shaoqing Ren}, {and} \bibinfo{person}{Jian Sun}.}
  \bibinfo{year}{2016}\natexlab{}.
\newblock \showarticletitle{Deep residual learning for image recognition}. In
  \bibinfo{booktitle}{\emph{CVPR}}. \bibinfo{pages}{770--778}.
\newblock


\bibitem[He et~al\mbox{.}(2020)]%
        {he2020lightgcn}
\bibfield{author}{\bibinfo{person}{Xiangnan He}, \bibinfo{person}{Kuan Deng},
  \bibinfo{person}{Xiang Wang}, \bibinfo{person}{Yan Li},
  \bibinfo{person}{Yongdong Zhang}, {and} \bibinfo{person}{Meng Wang}.}
  \bibinfo{year}{2020}\natexlab{}.
\newblock \showarticletitle{Lightgcn: Simplifying and powering graph
  convolution network for recommendation}. In
  \bibinfo{booktitle}{\emph{SIGIR}}. \bibinfo{pages}{639--648}.
\newblock


\bibitem[He et~al\mbox{.}(2017)]%
        {he2017neural}
\bibfield{author}{\bibinfo{person}{Xiangnan He}, \bibinfo{person}{Lizi Liao},
  \bibinfo{person}{Hanwang Zhang}, \bibinfo{person}{Liqiang Nie},
  \bibinfo{person}{Xia Hu}, {and} \bibinfo{person}{Tat-Seng Chua}.}
  \bibinfo{year}{2017}\natexlab{}.
\newblock \showarticletitle{Neural collaborative filtering}. In
  \bibinfo{booktitle}{\emph{TheWebConf}}. \bibinfo{pages}{173--182}.
\newblock


\bibitem[Huang et~al\mbox{.}(2021)]%
        {huang2021coupled}
\bibfield{author}{\bibinfo{person}{Zijie Huang}, \bibinfo{person}{Yizhou Sun},
  {and} \bibinfo{person}{Wei Wang}.} \bibinfo{year}{2021}\natexlab{}.
\newblock \showarticletitle{Coupled graph ode for learning interacting system
  dynamics}. In \bibinfo{booktitle}{\emph{KDD}}.
\newblock


\bibitem[Hui et~al\mbox{.}(2021)]%
        {hui2021trajnet}
\bibfield{author}{\bibinfo{person}{Bo Hui}, \bibinfo{person}{Da Yan},
  \bibinfo{person}{Haiquan Chen}, {and} \bibinfo{person}{Wei-Shinn Ku}.}
  \bibinfo{year}{2021}\natexlab{}.
\newblock \showarticletitle{Trajnet: A trajectory-based deep learning model for
  traffic prediction}. In \bibinfo{booktitle}{\emph{KDD}}.
  \bibinfo{pages}{716--724}.
\newblock


\bibitem[Jannach and Ludewig(2017)]%
        {jannach2017recurrent}
\bibfield{author}{\bibinfo{person}{Dietmar Jannach} {and}
  \bibinfo{person}{Malte Ludewig}.} \bibinfo{year}{2017}\natexlab{}.
\newblock \showarticletitle{When recurrent neural networks meet the
  neighborhood for session-based recommendation}. In
  \bibinfo{booktitle}{\emph{RecSys}}. \bibinfo{pages}{306--310}.
\newblock


\bibitem[Java et~al\mbox{.}(2009)]%
        {java2009we}
\bibfield{author}{\bibinfo{person}{Akshay Java}, \bibinfo{person}{Xiaodan
  Song}, \bibinfo{person}{Tim Finin}, {and} \bibinfo{person}{Belle Tseng}.}
  \bibinfo{year}{2009}\natexlab{}.
\newblock \showarticletitle{Why we twitter: An analysis of a microblogging
  community}. In \bibinfo{booktitle}{\emph{Advances in Web Mining and Web Usage
  Analysis: 9th International Workshop on Knowledge Discovery on the Web,
  WebKDD 2007, and 1st International Workshop on Social Networks Analysis,
  SNA-KDD 2007, San Jose, CA, USA, August 12-15, 2007. Revised Papers}}.
  \bibinfo{pages}{118--138}.
\newblock


\bibitem[Jenders et~al\mbox{.}(2013)]%
        {jenders2013analyzing}
\bibfield{author}{\bibinfo{person}{Maximilian Jenders},
  \bibinfo{person}{Gjergji Kasneci}, {and} \bibinfo{person}{Felix Naumann}.}
  \bibinfo{year}{2013}\natexlab{}.
\newblock \showarticletitle{Analyzing and predicting viral tweets}. In
  \bibinfo{booktitle}{\emph{TheWebConf}}. \bibinfo{pages}{657--664}.
\newblock


\bibitem[Jiang et~al\mbox{.}(2020)]%
        {jiang2020clicking}
\bibfield{author}{\bibinfo{person}{Guoyin Jiang}, \bibinfo{person}{Xiaodong
  Feng}, \bibinfo{person}{Wenping Liu}, {and} \bibinfo{person}{Xingjun Liu}.}
  \bibinfo{year}{2020}\natexlab{}.
\newblock \showarticletitle{Clicking position and user posting behavior in
  online review systems: A data-driven agent-based modeling approach}.
\newblock \bibinfo{journal}{\emph{Information Sciences}}  \bibinfo{volume}{512}
  (\bibinfo{year}{2020}), \bibinfo{pages}{161--174}.
\newblock


\bibitem[Jin et~al\mbox{.}(2023a)]%
        {jin2022code}
\bibfield{author}{\bibinfo{person}{Yiqiao Jin}, \bibinfo{person}{Yunsheng Bai},
  \bibinfo{person}{Yanqiao Zhu}, \bibinfo{person}{Yizhou Sun}, {and}
  \bibinfo{person}{Wei Wang}.} \bibinfo{year}{2023}\natexlab{a}.
\newblock \showarticletitle{Code Recommendation for Open Source Software
  Developers}. In \bibinfo{booktitle}{\emph{WWW}}.
\newblock


\bibitem[Jin et~al\mbox{.}(2023b)]%
        {10.1145/3543507.3583503}
\bibfield{author}{\bibinfo{person}{Yiqiao Jin}, \bibinfo{person}{Yunsheng Bai},
  \bibinfo{person}{Yanqiao Zhu}, \bibinfo{person}{Yizhou Sun}, {and}
  \bibinfo{person}{Wei Wang}.} \bibinfo{year}{2023}\natexlab{b}.
\newblock \showarticletitle{Code Recommendation for Open Source Software
  Developers}. In \bibinfo{booktitle}{\emph{TheWebConf}}.
  \bibinfo{pages}{1324–1333}.
\newblock


\bibitem[Jin et~al\mbox{.}(2023c)]%
        {jin2022prototypical}
\bibfield{author}{\bibinfo{person}{Yiqiao Jin}, \bibinfo{person}{Xiting Wang},
  \bibinfo{person}{Yaru Hao}, \bibinfo{person}{Yizhou Sun}, {and}
  \bibinfo{person}{Xing Xie}.} \bibinfo{year}{2023}\natexlab{c}.
\newblock \showarticletitle{Prototypical Fine-tuning: Towards Robust
  Performance Under Varying Data Sizes}.
\newblock \bibinfo{journal}{\emph{AAAI}} (\bibinfo{year}{2023}).
\newblock


\bibitem[Jin et~al\mbox{.}(2022)]%
        {jin2022towards}
\bibfield{author}{\bibinfo{person}{Yiqiao Jin}, \bibinfo{person}{Xiting Wang},
  \bibinfo{person}{Ruichao Yang}, \bibinfo{person}{Yizhou Sun},
  \bibinfo{person}{Wei Wang}, \bibinfo{person}{Hao Liao}, {and}
  \bibinfo{person}{Xing Xie}.} \bibinfo{year}{2022}\natexlab{}.
\newblock \showarticletitle{Towards fine-grained reasoning for fake news
  detection}. In \bibinfo{booktitle}{\emph{AAAI}}, Vol.~\bibinfo{volume}{36}.
  \bibinfo{pages}{5746--5754}.
\newblock


\bibitem[Jung et~al\mbox{.}(2021)]%
        {jung2021learning}
\bibfield{author}{\bibinfo{person}{Jaehun Jung}, \bibinfo{person}{Jinhong
  Jung}, {and} \bibinfo{person}{U Kang}.} \bibinfo{year}{2021}\natexlab{}.
\newblock \showarticletitle{Learning to walk across time for interpretable
  temporal knowledge graph completion}. In \bibinfo{booktitle}{\emph{KDD}}.
  \bibinfo{pages}{786--795}.
\newblock


\bibitem[Kazemi et~al\mbox{.}(2020)]%
        {kazemi2020representation}
\bibfield{author}{\bibinfo{person}{Seyed~Mehran Kazemi},
  \bibinfo{person}{Rishab Goel}, \bibinfo{person}{Kshitij Jain},
  \bibinfo{person}{Ivan Kobyzev}, \bibinfo{person}{Akshay Sethi},
  \bibinfo{person}{Peter Forsyth}, {and} \bibinfo{person}{Pascal Poupart}.}
  \bibinfo{year}{2020}\natexlab{}.
\newblock \showarticletitle{Representation learning for dynamic graphs: A
  survey}.
\newblock \bibinfo{journal}{\emph{JMLR}} \bibinfo{volume}{21},
  \bibinfo{number}{1} (\bibinfo{year}{2020}), \bibinfo{pages}{2648--2720}.
\newblock


\bibitem[Keegan(2019)]%
        {keegan2019dynamics}
\bibfield{author}{\bibinfo{person}{Brian~C Keegan}.}
  \bibinfo{year}{2019}\natexlab{}.
\newblock \showarticletitle{The Dynamics of Peer-Produced Political Information
  During the 2016 US Presidential Campaign}.
\newblock \bibinfo{journal}{\emph{HCI}} \bibinfo{volume}{3},
  \bibinfo{number}{CSCW} (\bibinfo{year}{2019}), \bibinfo{pages}{1--20}.
\newblock


\bibitem[Kim et~al\mbox{.}(2022)]%
        {KimLSK22}
\bibfield{author}{\bibinfo{person}{Taeri Kim}, \bibinfo{person}{Yeon{-}Chang
  Lee}, \bibinfo{person}{Kijung Shin}, {and} \bibinfo{person}{Sang{-}Wook
  Kim}.} \bibinfo{year}{2022}\natexlab{}.
\newblock \showarticletitle{{MARIO:} Modality-Aware Attention and
  Modality-Preserving Decoders for Multimedia Recommendation}. In
  \bibinfo{booktitle}{\emph{CIKM}}. \bibinfo{pages}{993--1002}.
\newblock


\bibitem[Kingma and Ba(2015)]%
        {kingma2015adam}
\bibfield{author}{\bibinfo{person}{Diederik~P Kingma} {and}
  \bibinfo{person}{Jimmy Ba}.} \bibinfo{year}{2015}\natexlab{}.
\newblock \showarticletitle{Adam: A Method for Stochastic Optimization}. In
  \bibinfo{booktitle}{\emph{ICLR}}.
\newblock


\bibitem[Kipf and Welling(2016)]%
        {kipf2016semi}
\bibfield{author}{\bibinfo{person}{Thomas~N Kipf} {and} \bibinfo{person}{Max
  Welling}.} \bibinfo{year}{2016}\natexlab{}.
\newblock \showarticletitle{Semi-supervised classification with graph
  convolutional networks}. In \bibinfo{booktitle}{\emph{ICLR}}.
\newblock


\bibitem[Kong et~al\mbox{.}(2022)]%
        {KongKJ0LPK22}
\bibfield{author}{\bibinfo{person}{Taeyong Kong}, \bibinfo{person}{Taeri Kim},
  \bibinfo{person}{Jinsung Jeon}, \bibinfo{person}{Jeongwhan Choi},
  \bibinfo{person}{Yeon{-}Chang Lee}, \bibinfo{person}{Noseong Park}, {and}
  \bibinfo{person}{Sang{-}Wook Kim}.} \bibinfo{year}{2022}\natexlab{}.
\newblock \showarticletitle{Linear, or Non-Linear, That is the Question!}. In
  \bibinfo{booktitle}{\emph{WSDM}}. \bibinfo{pages}{517--525}.
\newblock


\bibitem[Kumar et~al\mbox{.}(2018)]%
        {kumar2018community}
\bibfield{author}{\bibinfo{person}{Srijan Kumar}, \bibinfo{person}{William~L
  Hamilton}, \bibinfo{person}{Jure Leskovec}, {and} \bibinfo{person}{Dan
  Jurafsky}.} \bibinfo{year}{2018}\natexlab{}.
\newblock \showarticletitle{Community interaction and conflict on the web}. In
  \bibinfo{booktitle}{\emph{TheWebConf}}. \bibinfo{pages}{933--943}.
\newblock


\bibitem[Kumar et~al\mbox{.}(2019)]%
        {kumar2019predicting}
\bibfield{author}{\bibinfo{person}{Srijan Kumar}, \bibinfo{person}{Xikun
  Zhang}, {and} \bibinfo{person}{Jure Leskovec}.}
  \bibinfo{year}{2019}\natexlab{}.
\newblock \showarticletitle{Predicting dynamic embedding trajectory in temporal
  interaction networks}. In \bibinfo{booktitle}{\emph{KDD}}.
  \bibinfo{pages}{1269--1278}.
\newblock


\bibitem[Kwak et~al\mbox{.}(2010)]%
        {kwak2010twitter}
\bibfield{author}{\bibinfo{person}{Haewoon Kwak}, \bibinfo{person}{Changhyun
  Lee}, \bibinfo{person}{Hosung Park}, {and} \bibinfo{person}{Sue Moon}.}
  \bibinfo{year}{2010}\natexlab{}.
\newblock \showarticletitle{What is Twitter, a social network or a news
  media?}. In \bibinfo{booktitle}{\emph{TheWebConf}}.
  \bibinfo{pages}{591--600}.
\newblock


\bibitem[LeCun et~al\mbox{.}(1995)]%
        {lecun1995convolutional}
\bibfield{author}{\bibinfo{person}{Yann LeCun}, \bibinfo{person}{Yoshua
  Bengio}, {et~al\mbox{.}}} \bibinfo{year}{1995}\natexlab{}.
\newblock \showarticletitle{Convolutional networks for images, speech, and time
  series}.
\newblock \bibinfo{journal}{\emph{The Handbook of Brain Theory and Neural
  Networks}} \bibinfo{volume}{3361}, \bibinfo{number}{10}
  (\bibinfo{year}{1995}), \bibinfo{pages}{1995}.
\newblock


\bibitem[Lee et~al\mbox{.}(2022)]%
        {LeeL0K22}
\bibfield{author}{\bibinfo{person}{Yeon{-}Chang Lee}, \bibinfo{person}{JaeHyun
  Lee}, \bibinfo{person}{Dongwon Lee}, {and} \bibinfo{person}{Sang{-}Wook
  Kim}.} \bibinfo{year}{2022}\natexlab{}.
\newblock \showarticletitle{{THOR:} Self-Supervised Temporal Knowledge Graph
  Embedding via Three-Tower Graph Convolutional Networks}. In
  \bibinfo{booktitle}{\emph{ICDM}}. \bibinfo{pages}{1035--1040}.
\newblock


\bibitem[Leskovec et~al\mbox{.}(2009)]%
        {leskovec2009meme}
\bibfield{author}{\bibinfo{person}{Jure Leskovec}, \bibinfo{person}{Lars
  Backstrom}, {and} \bibinfo{person}{Jon Kleinberg}.}
  \bibinfo{year}{2009}\natexlab{}.
\newblock \showarticletitle{Meme-tracking and the dynamics of the news cycle}.
  In \bibinfo{booktitle}{\emph{KDD}}. \bibinfo{pages}{497--506}.
\newblock


\bibitem[Leung et~al\mbox{.}(2019)]%
        {leung2019personalized}
\bibfield{author}{\bibinfo{person}{Carson~K Leung}, \bibinfo{person}{Alfredo
  Cuzzocrea}, \bibinfo{person}{Jiaxing~Jason Mai}, \bibinfo{person}{Deyu Deng},
  {and} \bibinfo{person}{Fan Jiang}.} \bibinfo{year}{2019}\natexlab{}.
\newblock \showarticletitle{Personalized DeepInf: enhanced social influence
  prediction with deep learning and transfer learning}. In
  \bibinfo{booktitle}{\emph{BigData}}. \bibinfo{pages}{2871--2880}.
\newblock


\bibitem[Li et~al\mbox{.}(2017a)]%
        {li2017deepcas}
\bibfield{author}{\bibinfo{person}{Cheng Li}, \bibinfo{person}{Jiaqi Ma},
  \bibinfo{person}{Xiaoxiao Guo}, {and} \bibinfo{person}{Qiaozhu Mei}.}
  \bibinfo{year}{2017}\natexlab{a}.
\newblock \showarticletitle{Deepcas: An end-to-end predictor of information
  cascades}. In \bibinfo{booktitle}{\emph{TheWebConf}}.
  \bibinfo{pages}{577--586}.
\newblock


\bibitem[Li et~al\mbox{.}(2013)]%
        {li2013popularity}
\bibfield{author}{\bibinfo{person}{Haitao Li}, \bibinfo{person}{Xiaoqiang Ma},
  \bibinfo{person}{Feng Wang}, \bibinfo{person}{Jiangchuan Liu}, {and}
  \bibinfo{person}{Ke Xu}.} \bibinfo{year}{2013}\natexlab{}.
\newblock \showarticletitle{On popularity prediction of videos shared in online
  social networks}. In \bibinfo{booktitle}{\emph{CIKM}}.
  \bibinfo{pages}{169--178}.
\newblock


\bibitem[Li et~al\mbox{.}(2017b)]%
        {li2017neural}
\bibfield{author}{\bibinfo{person}{Jing Li}, \bibinfo{person}{Pengjie Ren},
  \bibinfo{person}{Zhumin Chen}, \bibinfo{person}{Zhaochun Ren},
  \bibinfo{person}{Tao Lian}, {and} \bibinfo{person}{Jun Ma}.}
  \bibinfo{year}{2017}\natexlab{b}.
\newblock \showarticletitle{Neural attentive session-based recommendation}. In
  \bibinfo{booktitle}{\emph{CIKM}}. \bibinfo{pages}{1419--1428}.
\newblock


\bibitem[Li et~al\mbox{.}(2020)]%
        {li2020time}
\bibfield{author}{\bibinfo{person}{Jiacheng Li}, \bibinfo{person}{Yujie Wang},
  {and} \bibinfo{person}{Julian McAuley}.} \bibinfo{year}{2020}\natexlab{}.
\newblock \showarticletitle{Time interval aware self-attention for sequential
  recommendation}. In \bibinfo{booktitle}{\emph{WSDM}}.
  \bibinfo{pages}{322--330}.
\newblock


\bibitem[Lin et~al\mbox{.}(2017)]%
        {lin2017better}
\bibfield{author}{\bibinfo{person}{Zhiyuan Lin}, \bibinfo{person}{Niloufar
  Salehi}, \bibinfo{person}{Bowen Yao}, \bibinfo{person}{Yiqi Chen}, {and}
  \bibinfo{person}{Michael~S Bernstein}.} \bibinfo{year}{2017}\natexlab{}.
\newblock \showarticletitle{Better when it was smaller? community content and
  behavior after massive growth}. In \bibinfo{booktitle}{\emph{ICWSM}}.
\newblock


\bibitem[Ling et~al\mbox{.}(2021)]%
        {ling2021dissecting}
\bibfield{author}{\bibinfo{person}{Chen Ling}, \bibinfo{person}{Ihab AbuHilal},
  \bibinfo{person}{Jeremy Blackburn}, \bibinfo{person}{Emiliano De~Cristofaro},
  \bibinfo{person}{Savvas Zannettou}, {and} \bibinfo{person}{Gianluca
  Stringhini}.} \bibinfo{year}{2021}\natexlab{}.
\newblock \showarticletitle{Dissecting the meme magic: Understanding indicators
  of virality in image memes}.
\newblock \bibinfo{journal}{\emph{ACM HCI}} \bibinfo{volume}{5},
  \bibinfo{number}{CSCW1} (\bibinfo{year}{2021}), \bibinfo{pages}{1--24}.
\newblock


\bibitem[Liu et~al\mbox{.}(2023a)]%
        {S2T_ML}
\bibfield{author}{\bibinfo{person}{Meng Liu}, \bibinfo{person}{Ke Liang},
  \bibinfo{person}{Bin Xiao}, \bibinfo{person}{Sihang Zhou},
  \bibinfo{person}{Wenxuan Tu}, \bibinfo{person}{Yue Liu},
  \bibinfo{person}{Xihong Yang}, {and} \bibinfo{person}{Xinwang Liu}.}
  \bibinfo{year}{2023}\natexlab{a}.
\newblock \showarticletitle{Self-Supervised Temporal Graph learning with
  Temporal and Structural Intensity Alignment}.
\newblock \bibinfo{journal}{\emph{arXiv:2302.07491}} (\bibinfo{year}{2023}).
\newblock


\bibitem[Liu and Liu(2021)]%
        {MNCI_ML_SIGIR}
\bibfield{author}{\bibinfo{person}{Meng Liu} {and} \bibinfo{person}{Yong Liu}.}
  \bibinfo{year}{2021}\natexlab{}.
\newblock \showarticletitle{Inductive representation learning in temporal
  networks via mining neighborhood and community influences}. In
  \bibinfo{booktitle}{\emph{SIGIR}}. \bibinfo{pages}{2202--2206}.
\newblock


\bibitem[Liu et~al\mbox{.}(2023b)]%
        {TGC_ML}
\bibfield{author}{\bibinfo{person}{Meng Liu}, \bibinfo{person}{Yue Liu},
  \bibinfo{person}{Ke Liang}, \bibinfo{person}{Siwei Wang},
  \bibinfo{person}{Sihang Zhou}, {and} \bibinfo{person}{Xinwang Liu}.}
  \bibinfo{year}{2023}\natexlab{b}.
\newblock \showarticletitle{Deep Temporal Graph Clustering}.
\newblock \bibinfo{journal}{\emph{arXiv:2305.10738}} (\bibinfo{year}{2023}).
\newblock


\bibitem[Logan et~al\mbox{.}(2000)]%
        {logan2000mel}
\bibfield{author}{\bibinfo{person}{Beth Logan} {et~al\mbox{.}}}
  \bibinfo{year}{2000}\natexlab{}.
\newblock \showarticletitle{Mel frequency cepstral coefficients for music
  modeling.}. In \bibinfo{booktitle}{\emph{ISMIR}}, Vol.~\bibinfo{volume}{270}.
  \bibinfo{pages}{11}.
\newblock


\bibitem[Ludewig and Jannach(2018)]%
        {ludewig2018evaluation}
\bibfield{author}{\bibinfo{person}{Malte Ludewig} {and}
  \bibinfo{person}{Dietmar Jannach}.} \bibinfo{year}{2018}\natexlab{}.
\newblock \showarticletitle{Evaluation of session-based recommendation
  algorithms}.
\newblock \bibinfo{journal}{\emph{User Modeling and User-Adapted Interaction}}
  \bibinfo{volume}{28} (\bibinfo{year}{2018}), \bibinfo{pages}{331--390}.
\newblock


\bibitem[Mao et~al\mbox{.}(2022)]%
        {mao2022whole}
\bibfield{author}{\bibinfo{person}{Haitao Mao}, \bibinfo{person}{Lixin Zou},
  \bibinfo{person}{Yujia Zheng}, \bibinfo{person}{Jiliang Tang},
  \bibinfo{person}{Xiaokai Chu}, \bibinfo{person}{Jiashu Zhao}, {and}
  \bibinfo{person}{Dawei Yin}.} \bibinfo{year}{2022}\natexlab{}.
\newblock \showarticletitle{Whole Page Unbiased Learning to Rank}.
\newblock \bibinfo{journal}{\emph{arXiv preprint arXiv:2210.10718}}
  (\bibinfo{year}{2022}).
\newblock


\bibitem[Meng et~al\mbox{.}(2023)]%
        {meng2023recognize}
\bibfield{author}{\bibinfo{person}{Qing Meng}, \bibinfo{person}{Hui Yan},
  \bibinfo{person}{Bo Liu}, \bibinfo{person}{Xiangguo Sun},
  \bibinfo{person}{Mingrui Hu}, {and} \bibinfo{person}{Jiuxin Cao}.}
  \bibinfo{year}{2023}\natexlab{}.
\newblock \showarticletitle{Recognize News Transition from Collective Behavior
  for News Recommendation}.
\newblock \bibinfo{journal}{\emph{TOIS}} \bibinfo{volume}{41},
  \bibinfo{number}{4} (\bibinfo{year}{2023}), \bibinfo{pages}{1--30}.
\newblock


\bibitem[Micallef et~al\mbox{.}(2020a)]%
        {micallef2020role}
\bibfield{author}{\bibinfo{person}{Nicholas Micallef}, \bibinfo{person}{Bing
  He}, \bibinfo{person}{Srijan Kumar}, \bibinfo{person}{Mustaque Ahamad}, {and}
  \bibinfo{person}{Nasir Memon}.} \bibinfo{year}{2020}\natexlab{a}.
\newblock \showarticletitle{The role of the crowd in countering misinformation:
  A case study of the COVID-19 infodemic}. In
  \bibinfo{booktitle}{\emph{BigData}}. \bibinfo{pages}{748--757}.
\newblock


\bibitem[Micallef et~al\mbox{.}(2020b)]%
        {Micallef2020TheRO}
\bibfield{author}{\bibinfo{person}{Nicholas Micallef}, \bibinfo{person}{Bing
  He}, \bibinfo{person}{Srijan Kumar}, \bibinfo{person}{Mustaque Ahamad}, {and}
  \bibinfo{person}{Nasir~D. Memon}.} \bibinfo{year}{2020}\natexlab{b}.
\newblock \showarticletitle{The Role of the Crowd in Countering Misinformation:
  A Case Study of the COVID-19 Infodemic}.
\newblock \bibinfo{journal}{\emph{BigData}} (\bibinfo{year}{2020}),
  \bibinfo{pages}{748--757}.
\newblock


\bibitem[Micallef et~al\mbox{.}(2022)]%
        {Micallef2022CrossPlatformMM}
\bibfield{author}{\bibinfo{person}{Nicholas Micallef}, \bibinfo{person}{Marcelo
  Sandoval-Casta{\~n}eda}, \bibinfo{person}{Adir Cohen},
  \bibinfo{person}{Mustaque Ahamad}, \bibinfo{person}{Srijan},
  \bibinfo{person}{Kumar}, {and} \bibinfo{person}{Nasir~D. Memon}.}
  \bibinfo{year}{2022}\natexlab{}.
\newblock \showarticletitle{Cross-Platform Multimodal Misinformation: Taxonomy,
  Characteristics and Detection for Textual Posts and Videos}. In
  \bibinfo{booktitle}{\emph{ICWSM}}.
\newblock


\bibitem[Myers and Leskovec(2014)]%
        {myers2014bursty}
\bibfield{author}{\bibinfo{person}{Seth~A Myers} {and} \bibinfo{person}{Jure
  Leskovec}.} \bibinfo{year}{2014}\natexlab{}.
\newblock \showarticletitle{The bursty dynamics of the twitter information
  network}. In \bibinfo{booktitle}{\emph{TheWebConf}}.
  \bibinfo{pages}{913--924}.
\newblock


\bibitem[Naseem et~al\mbox{.}(2022)]%
        {naseem2022early}
\bibfield{author}{\bibinfo{person}{Usman Naseem}, \bibinfo{person}{Adam~G
  Dunn}, \bibinfo{person}{Jinman Kim}, {and} \bibinfo{person}{Matloob Khushi}.}
  \bibinfo{year}{2022}\natexlab{}.
\newblock \showarticletitle{Early identification of depression severity levels
  on reddit using ordinal classification}. In
  \bibinfo{booktitle}{\emph{TheWebConf}}. \bibinfo{pages}{2563--2572}.
\newblock


\bibitem[Ng et~al\mbox{.}(2022)]%
        {ng2022cross}
\bibfield{author}{\bibinfo{person}{Lynnette Hui~Xian Ng},
  \bibinfo{person}{Iain~J Cruickshank}, {and} \bibinfo{person}{Kathleen~M
  Carley}.} \bibinfo{year}{2022}\natexlab{}.
\newblock \showarticletitle{Cross-platform information spread during the
  january 6th capitol riots}.
\newblock \bibinfo{journal}{\emph{Social Network Analysis and Mining}}
  \bibinfo{volume}{12}, \bibinfo{number}{1} (\bibinfo{year}{2022}),
  \bibinfo{pages}{133}.
\newblock


\bibitem[Oh et~al\mbox{.}(2022)]%
        {oh2022implicit}
\bibfield{author}{\bibinfo{person}{Sejoon Oh}, \bibinfo{person}{Ankur
  Bhardwaj}, \bibinfo{person}{Jongseok Han}, \bibinfo{person}{Sungchul Kim},
  \bibinfo{person}{Ryan~A Rossi}, {and} \bibinfo{person}{Srijan Kumar}.}
  \bibinfo{year}{2022}\natexlab{}.
\newblock \showarticletitle{Implicit Session Contexts for Next-Item
  Recommendations}. In \bibinfo{booktitle}{\emph{CIKM}}.
  \bibinfo{pages}{4364--4368}.
\newblock


\bibitem[Okawa and Iwata(2022)]%
        {Okawa2022PredictingOD}
\bibfield{author}{\bibinfo{person}{Maya Okawa} {and} \bibinfo{person}{Tomoharu
  Iwata}.} \bibinfo{year}{2022}\natexlab{}.
\newblock \showarticletitle{Predicting Opinion Dynamics via
  Sociologically-Informed Neural Networks}.
\newblock \bibinfo{journal}{\emph{KDD}} (\bibinfo{year}{2022}).
\newblock


\bibitem[Pacheco et~al\mbox{.}(2021)]%
        {pacheco2021uncovering}
\bibfield{author}{\bibinfo{person}{Diogo Pacheco}, \bibinfo{person}{Pik-Mai
  Hui}, \bibinfo{person}{Christopher Torres-Lugo}, \bibinfo{person}{Bao~Tran
  Truong}, \bibinfo{person}{Alessandro Flammini}, {and}
  \bibinfo{person}{Filippo Menczer}.} \bibinfo{year}{2021}\natexlab{}.
\newblock \showarticletitle{Uncovering Coordinated Networks on Social Media:
  Methods and Case Studies}.
\newblock \bibinfo{journal}{\emph{ICWSM}}  \bibinfo{volume}{21}
  (\bibinfo{year}{2021}), \bibinfo{pages}{455--466}.
\newblock


\bibitem[Paszke et~al\mbox{.}(2019)]%
        {paszke2019pytorch}
\bibfield{author}{\bibinfo{person}{Adam Paszke}, \bibinfo{person}{Sam Gross},
  \bibinfo{person}{Francisco Massa}, \bibinfo{person}{Adam Lerer},
  \bibinfo{person}{James Bradbury}, \bibinfo{person}{Gregory Chanan},
  \bibinfo{person}{Trevor Killeen}, \bibinfo{person}{Zeming Lin},
  \bibinfo{person}{Natalia Gimelshein}, \bibinfo{person}{Luca Antiga},
  {et~al\mbox{.}}} \bibinfo{year}{2019}\natexlab{}.
\newblock \showarticletitle{Pytorch: An imperative style, high-performance deep
  learning library}.
\newblock \bibinfo{journal}{\emph{NIPS}}  \bibinfo{volume}{32}.
\newblock


\bibitem[Peng et~al\mbox{.}(2022)]%
        {peng2022svd}
\bibfield{author}{\bibinfo{person}{Shaowen Peng}, \bibinfo{person}{Kazunari
  Sugiyama}, {and} \bibinfo{person}{Tsunenori Mine}.}
  \bibinfo{year}{2022}\natexlab{}.
\newblock \showarticletitle{SVD-GCN: A Simplified Graph Convolution Paradigm
  for Recommendation}. In \bibinfo{booktitle}{\emph{CIKM}}.
  \bibinfo{pages}{1625--1634}.
\newblock


\bibitem[Phadke et~al\mbox{.}(2021)]%
        {phadke2021makes}
\bibfield{author}{\bibinfo{person}{Shruti Phadke}, \bibinfo{person}{Mattia
  Samory}, {and} \bibinfo{person}{Tanushree Mitra}.}
  \bibinfo{year}{2021}\natexlab{}.
\newblock \showarticletitle{What makes people join conspiracy communities? Role
  of social factors in conspiracy engagement}.
\newblock \bibinfo{journal}{\emph{HCI}} \bibinfo{volume}{4},
  \bibinfo{number}{CSCW3} (\bibinfo{year}{2021}), \bibinfo{pages}{1--30}.
\newblock


\bibitem[Qiu et~al\mbox{.}(2018)]%
        {qiu2018deepinf}
\bibfield{author}{\bibinfo{person}{Jiezhong Qiu}, \bibinfo{person}{Jian Tang},
  \bibinfo{person}{Hao Ma}, \bibinfo{person}{Yuxiao Dong},
  \bibinfo{person}{Kuansan Wang}, {and} \bibinfo{person}{Jie Tang}.}
  \bibinfo{year}{2018}\natexlab{}.
\newblock \showarticletitle{Deepinf: Social influence prediction with deep
  learning}. In \bibinfo{booktitle}{\emph{CIKM}}. \bibinfo{pages}{2110--2119}.
\newblock


\bibitem[Qu et~al\mbox{.}(2022)]%
        {qu2022evolution}
\bibfield{author}{\bibinfo{person}{Yiting Qu}, \bibinfo{person}{Xinlei He},
  \bibinfo{person}{Shannon Pierson}, \bibinfo{person}{Michael Backes},
  \bibinfo{person}{Yang Zhang}, {and} \bibinfo{person}{Savvas Zannettou}.}
  \bibinfo{year}{2022}\natexlab{}.
\newblock \showarticletitle{On the Evolution of (Hateful) Memes by Means of
  Multimodal Contrastive Learning}. In \bibinfo{booktitle}{\emph{2023 IEEE
  Symposium on Security and Privacy (SP)}}. \bibinfo{pages}{1348--1365}.
\newblock


\bibitem[Rendle et~al\mbox{.}(2009)]%
        {rendle2009bpr}
\bibfield{author}{\bibinfo{person}{Steffen Rendle}, \bibinfo{person}{Christoph
  Freudenthaler}, \bibinfo{person}{Zeno Gantner}, {and} \bibinfo{person}{Lars
  Schmidt-Thieme}.} \bibinfo{year}{2009}\natexlab{}.
\newblock \showarticletitle{BPR: Bayesian personalized ranking from implicit
  feedback}. In \bibinfo{booktitle}{\emph{UAI}}. \bibinfo{pages}{452--461}.
\newblock


\bibitem[Rong et~al\mbox{.}(2016)]%
        {rong2016model}
\bibfield{author}{\bibinfo{person}{Yu Rong}, \bibinfo{person}{Qiankun Zhu},
  {and} \bibinfo{person}{Hong Cheng}.} \bibinfo{year}{2016}\natexlab{}.
\newblock \showarticletitle{A model-free approach to infer the diffusion
  network from event cascade}. In \bibinfo{booktitle}{\emph{CIKM}}.
  \bibinfo{pages}{1653--1662}.
\newblock


\bibitem[Rossi et~al\mbox{.}(2020)]%
        {rossi2020temporal}
\bibfield{author}{\bibinfo{person}{Emanuele Rossi}, \bibinfo{person}{Ben
  Chamberlain}, \bibinfo{person}{Fabrizio Frasca}, \bibinfo{person}{Davide
  Eynard}, \bibinfo{person}{Federico Monti}, {and} \bibinfo{person}{Michael
  Bronstein}.} \bibinfo{year}{2020}\natexlab{}.
\newblock \showarticletitle{Temporal graph networks for deep learning on
  dynamic graphs}.
\newblock \bibinfo{journal}{\emph{arXiv:2006.10637}} (\bibinfo{year}{2020}).
\newblock


\bibitem[Shi et~al\mbox{.}(2021)]%
        {shimasked}
\bibfield{author}{\bibinfo{person}{Yunsheng Shi}, \bibinfo{person}{Zhengjie
  Huang}, \bibinfo{person}{Shikun Feng}, \bibinfo{person}{Hui Zhong},
  \bibinfo{person}{Wenjing Wang}, {and} \bibinfo{person}{Yu Sun}.}
  \bibinfo{year}{2021}\natexlab{}.
\newblock \showarticletitle{Masked Label Prediction: Unified Message Passing
  Model for Semi-Supervised Classification}. In
  \bibinfo{booktitle}{\emph{IJCAI}}.
\newblock


\bibitem[Shin et~al\mbox{.}(2018)]%
        {shin2018diffusion}
\bibfield{author}{\bibinfo{person}{Jieun Shin}, \bibinfo{person}{Lian Jian},
  \bibinfo{person}{Kevin Driscoll}, {and} \bibinfo{person}{Fran{\c{c}}ois
  Bar}.} \bibinfo{year}{2018}\natexlab{}.
\newblock \showarticletitle{The diffusion of misinformation on social media:
  Temporal pattern, message, and source}.
\newblock \bibinfo{journal}{\emph{Computers in Human Behavior}}
  \bibinfo{volume}{83} (\bibinfo{year}{2018}), \bibinfo{pages}{278--287}.
\newblock


\bibitem[Shu et~al\mbox{.}(2019)]%
        {shu2019defend}
\bibfield{author}{\bibinfo{person}{Kai Shu}, \bibinfo{person}{Limeng Cui},
  \bibinfo{person}{Suhang Wang}, \bibinfo{person}{Dongwon Lee}, {and}
  \bibinfo{person}{Huan Liu}.} \bibinfo{year}{2019}\natexlab{}.
\newblock \showarticletitle{defend: Explainable fake news detection}. In
  \bibinfo{booktitle}{\emph{KDD}}. \bibinfo{pages}{395--405}.
\newblock


\bibitem[Song et~al\mbox{.}(2021)]%
        {song2021temporally}
\bibfield{author}{\bibinfo{person}{Chenguang Song}, \bibinfo{person}{Kai Shu},
  {and} \bibinfo{person}{Bin Wu}.} \bibinfo{year}{2021}\natexlab{}.
\newblock \showarticletitle{Temporally evolving graph neural network for fake
  news detection}.
\newblock \bibinfo{journal}{\emph{Information Processing \& Management}}
  \bibinfo{volume}{58}, \bibinfo{number}{6} (\bibinfo{year}{2021}),
  \bibinfo{pages}{102712}.
\newblock


\bibitem[Song et~al\mbox{.}(2022)]%
        {song2022friend}
\bibfield{author}{\bibinfo{person}{Xiran Song}, \bibinfo{person}{Jianxun Lian},
  \bibinfo{person}{Hong Huang}, \bibinfo{person}{Mingqi Wu},
  \bibinfo{person}{Hai Jin}, {and} \bibinfo{person}{Xing Xie}.}
  \bibinfo{year}{2022}\natexlab{}.
\newblock \showarticletitle{Friend Recommendations with Self-Rescaling Graph
  Neural Networks}. In \bibinfo{booktitle}{\emph{KDD}}.
  \bibinfo{pages}{3909--3919}.
\newblock


\bibitem[Statista(2021)]%
        {statista-social-networks}
\bibfield{author}{\bibinfo{person}{Statista}.} \bibinfo{year}{2021}\natexlab{}.
\newblock \bibinfo{booktitle}{\emph{Most popular social networks worldwide as
  of January 2023, ranked by number of monthly active users}}.
\newblock
\urldef\tempurl%
\url{https://www.statista.com/statistics/272014/global-social-networks-ranked-by-number-of-users/}
\showURL{%
\tempurl}


\bibitem[Sturm and Albrecht(2021)]%
        {sturm2021constituent}
\bibfield{author}{\bibinfo{person}{Tristan Sturm} {and} \bibinfo{person}{Tom
  Albrecht}.} \bibinfo{year}{2021}\natexlab{}.
\newblock \showarticletitle{Constituent Covid-19 apocalypses: contagious
  conspiracism, 5G, and viral vaccinations}.
\newblock \bibinfo{journal}{\emph{Anthropology \& medicine}}
  \bibinfo{volume}{28}, \bibinfo{number}{1} (\bibinfo{year}{2021}),
  \bibinfo{pages}{122--139}.
\newblock


\bibitem[Tan(2018)]%
        {tan2018tracing}
\bibfield{author}{\bibinfo{person}{Chenhao Tan}.}
  \bibinfo{year}{2018}\natexlab{}.
\newblock \showarticletitle{Tracing community genealogy: how new communities
  emerge from the old}. In \bibinfo{booktitle}{\emph{ICWSM}}.
\newblock


\bibitem[Veli{\v{c}}kovi{\'c} et~al\mbox{.}(2018)]%
        {velivckovic2018graph}
\bibfield{author}{\bibinfo{person}{Petar Veli{\v{c}}kovi{\'c}},
  \bibinfo{person}{Guillem Cucurull}, \bibinfo{person}{Arantxa Casanova},
  \bibinfo{person}{Adriana Romero}, \bibinfo{person}{Pietro Li{\`o}}, {and}
  \bibinfo{person}{Yoshua Bengio}.} \bibinfo{year}{2018}\natexlab{}.
\newblock \showarticletitle{Graph Attention Networks}. In
  \bibinfo{booktitle}{\emph{ICLR}}.
\newblock


\bibitem[Verma et~al\mbox{.}(2022)]%
        {verma2022examining}
\bibfield{author}{\bibinfo{person}{Gaurav Verma}, \bibinfo{person}{Ankur
  Bhardwaj}, \bibinfo{person}{Talayeh Aledavood}, \bibinfo{person}{Munmun
  De~Choudhury}, {and} \bibinfo{person}{Srijan Kumar}.}
  \bibinfo{year}{2022}\natexlab{}.
\newblock \showarticletitle{Examining the impact of sharing COVID-19
  misinformation online on mental health}.
\newblock \bibinfo{journal}{\emph{Scientific Reports}} \bibinfo{volume}{12},
  \bibinfo{number}{1} (\bibinfo{year}{2022}), \bibinfo{pages}{1--9}.
\newblock


\bibitem[Vosoughi et~al\mbox{.}(2018)]%
        {vosoughi2018spread}
\bibfield{author}{\bibinfo{person}{Soroush Vosoughi}, \bibinfo{person}{Deb
  Roy}, {and} \bibinfo{person}{Sinan Aral}.} \bibinfo{year}{2018}\natexlab{}.
\newblock \showarticletitle{The spread of true and false news online}.
\newblock \bibinfo{journal}{\emph{Science}} \bibinfo{volume}{359},
  \bibinfo{number}{6380} (\bibinfo{year}{2018}), \bibinfo{pages}{1146--1151}.
\newblock


\bibitem[Waller and Anderson(2019)]%
        {waller2019generalists}
\bibfield{author}{\bibinfo{person}{Isaac Waller} {and} \bibinfo{person}{Ashton
  Anderson}.} \bibinfo{year}{2019}\natexlab{}.
\newblock \showarticletitle{Generalists and specialists: Using community
  embeddings to quantify activity diversity in online platforms}. In
  \bibinfo{booktitle}{\emph{TheWebConf}}. \bibinfo{pages}{1954--1964}.
\newblock


\bibitem[Waller and Anderson(2021)]%
        {waller2021quantifying}
\bibfield{author}{\bibinfo{person}{Isaac Waller} {and} \bibinfo{person}{Ashton
  Anderson}.} \bibinfo{year}{2021}\natexlab{}.
\newblock \showarticletitle{Quantifying social organization and political
  polarization in online platforms}.
\newblock \bibinfo{journal}{\emph{Nature}} \bibinfo{volume}{600},
  \bibinfo{number}{7888} (\bibinfo{year}{2021}), \bibinfo{pages}{264--268}.
\newblock


\bibitem[Wang et~al\mbox{.}(2023a)]%
        {wang2023car}
\bibfield{author}{\bibinfo{person}{Weiqi Wang}, \bibinfo{person}{Tianqing
  Fang}, \bibinfo{person}{Wenxuan Ding}, \bibinfo{person}{Baixuan Xu},
  \bibinfo{person}{Xin Liu}, \bibinfo{person}{Yangqiu Song}, {and}
  \bibinfo{person}{Antoine Bosselut}.} \bibinfo{year}{2023}\natexlab{a}.
\newblock \showarticletitle{CAR: Conceptualization-Augmented Reasoner for
  Zero-Shot Commonsense Question Answering}.
\newblock \bibinfo{journal}{\emph{arXiv preprint arXiv:2305.14869}}
  (\bibinfo{year}{2023}).
\newblock


\bibitem[Wang et~al\mbox{.}(2023b)]%
        {wang2023cat}
\bibfield{author}{\bibinfo{person}{Weiqi Wang}, \bibinfo{person}{Tianqing
  Fang}, \bibinfo{person}{Baixuan Xu}, \bibinfo{person}{Chun Yi~Louis Bo},
  \bibinfo{person}{Yangqiu Song}, {and} \bibinfo{person}{Lei Chen}.}
  \bibinfo{year}{2023}\natexlab{b}.
\newblock \showarticletitle{CAT: A contextualized conceptualization and
  instantiation framework for commonsense reasoning}. In
  \bibinfo{booktitle}{\emph{ACL}}.
\newblock


\bibitem[Wang et~al\mbox{.}(2020)]%
        {wang2020minilm}
\bibfield{author}{\bibinfo{person}{Wenhui Wang}, \bibinfo{person}{Furu Wei},
  \bibinfo{person}{Li Dong}, \bibinfo{person}{Hangbo Bao}, \bibinfo{person}{Nan
  Yang}, {and} \bibinfo{person}{Ming Zhou}.} \bibinfo{year}{2020}\natexlab{}.
\newblock \showarticletitle{MiniLM: Deep self-attention distillation for
  task-agnostic compression of pre-trained transformers}.
\newblock \bibinfo{journal}{\emph{NIPS}}  \bibinfo{volume}{33}
  (\bibinfo{year}{2020}), \bibinfo{pages}{5776--5788}.
\newblock


\bibitem[Wang et~al\mbox{.}(2019)]%
        {wang2019neural}
\bibfield{author}{\bibinfo{person}{Xiang Wang}, \bibinfo{person}{Xiangnan He},
  \bibinfo{person}{Meng Wang}, \bibinfo{person}{Fuli Feng}, {and}
  \bibinfo{person}{Tat-Seng Chua}.} \bibinfo{year}{2019}\natexlab{}.
\newblock \showarticletitle{Neural graph collaborative filtering}. In
  \bibinfo{booktitle}{\emph{SIGIR}}. \bibinfo{pages}{165--174}.
\newblock


\bibitem[Wang et~al\mbox{.}(2022)]%
        {wang2022multi}
\bibfield{author}{\bibinfo{person}{Xiting Wang}, \bibinfo{person}{Kunpeng Liu},
  \bibinfo{person}{Dongjie Wang}, \bibinfo{person}{Le Wu},
  \bibinfo{person}{Yanjie Fu}, {and} \bibinfo{person}{Xing Xie}.}
  \bibinfo{year}{2022}\natexlab{}.
\newblock \showarticletitle{Multi-level recommendation reasoning over knowledge
  graphs with reinforcement learning}. In
  \bibinfo{booktitle}{\emph{TheWebConf}}. \bibinfo{pages}{2098--2108}.
\newblock


\bibitem[Wang et~al\mbox{.}(2021)]%
        {wang2021inductive}
\bibfield{author}{\bibinfo{person}{Yanbang Wang}, \bibinfo{person}{Yen-Yu
  Chang}, \bibinfo{person}{Yunyu Liu}, \bibinfo{person}{Jure Leskovec}, {and}
  \bibinfo{person}{Pan Li}.} \bibinfo{year}{2021}\natexlab{}.
\newblock \showarticletitle{Inductive representation learning in temporal
  networks via causal anonymous walks}.
\newblock \bibinfo{journal}{\emph{ICLR}} (\bibinfo{year}{2021}).
\newblock


\bibitem[Wikipedia(2023)]%
        {wiki-most-visited-website}
\bibfield{author}{\bibinfo{person}{Wikipedia}.}
  \bibinfo{year}{2023}\natexlab{}.
\newblock \bibinfo{booktitle}{\emph{List of Most Visited Websites}}.
\newblock
\newblock
\shownote{April 13, 2023}.


\bibitem[Wu et~al\mbox{.}(2022)]%
        {wu2022bias}
\bibfield{author}{\bibinfo{person}{Junfei Wu}, \bibinfo{person}{Qiang Liu},
  \bibinfo{person}{Weizhi Xu}, {and} \bibinfo{person}{Shu Wu}.}
  \bibinfo{year}{2022}\natexlab{}.
\newblock \showarticletitle{Bias mitigation for evidence-aware fake news
  detection by causal intervention}. In \bibinfo{booktitle}{\emph{SIGIR}}.
  \bibinfo{pages}{2308--2313}.
\newblock


\bibitem[Wu et~al\mbox{.}(2019)]%
        {wu2019session}
\bibfield{author}{\bibinfo{person}{Shu Wu}, \bibinfo{person}{Yuyuan Tang},
  \bibinfo{person}{Yanqiao Zhu}, \bibinfo{person}{Liang Wang},
  \bibinfo{person}{Xing Xie}, {and} \bibinfo{person}{Tieniu Tan}.}
  \bibinfo{year}{2019}\natexlab{}.
\newblock \showarticletitle{Session-based recommendation with graph neural
  networks}. In \bibinfo{booktitle}{\emph{AAAI}}, Vol.~\bibinfo{volume}{33}.
  \bibinfo{pages}{346--353}.
\newblock


\bibitem[Xia et~al\mbox{.}(2021)]%
        {xia2021deepis}
\bibfield{author}{\bibinfo{person}{Wenwen Xia}, \bibinfo{person}{Yuchen Li},
  \bibinfo{person}{Jun Wu}, {and} \bibinfo{person}{Shenghong Li}.}
  \bibinfo{year}{2021}\natexlab{}.
\newblock \showarticletitle{Deepis: Susceptibility estimation on social
  networks}. In \bibinfo{booktitle}{\emph{WSDM}}. \bibinfo{pages}{761--769}.
\newblock


\bibitem[Xu et~al\mbox{.}(2020)]%
        {xu2020inductive}
\bibfield{author}{\bibinfo{person}{Da Xu}, \bibinfo{person}{Chuanwei Ruan},
  \bibinfo{person}{Evren Korpeoglu}, \bibinfo{person}{Sushant Kumar}, {and}
  \bibinfo{person}{Kannan Achan}.} \bibinfo{year}{2020}\natexlab{}.
\newblock \showarticletitle{Inductive representation learning on temporal
  graphs}.
\newblock \bibinfo{journal}{\emph{ICLR}} (\bibinfo{year}{2020}).
\newblock


\bibitem[Xu et~al\mbox{.}(2018)]%
        {xu2018powerful}
\bibfield{author}{\bibinfo{person}{Keyulu Xu}, \bibinfo{person}{Weihua Hu},
  \bibinfo{person}{Jure Leskovec}, {and} \bibinfo{person}{Stefanie Jegelka}.}
  \bibinfo{year}{2018}\natexlab{}.
\newblock \showarticletitle{How Powerful are Graph Neural Networks?}. In
  \bibinfo{booktitle}{\emph{ICLR}}.
\newblock


\bibitem[Xu et~al\mbox{.}(2022a)]%
        {xu2022tervit}
\bibfield{author}{\bibinfo{person}{Sheng Xu}, \bibinfo{person}{Yanjing Li},
  \bibinfo{person}{Teli Ma}, \bibinfo{person}{Bohan Zeng},
  \bibinfo{person}{Baochang Zhang}, \bibinfo{person}{Peng Gao}, {and}
  \bibinfo{person}{Jinhu Lv}.} \bibinfo{year}{2022}\natexlab{a}.
\newblock \showarticletitle{TerViT: An efficient ternary vision transformer}.
\newblock \bibinfo{journal}{\emph{arXiv:2201.08050}} (\bibinfo{year}{2022}).
\newblock


\bibitem[Xu et~al\mbox{.}(2022b)]%
        {xu2022evidence}
\bibfield{author}{\bibinfo{person}{Weizhi Xu}, \bibinfo{person}{Junfei Wu},
  \bibinfo{person}{Qiang Liu}, \bibinfo{person}{Shu Wu}, {and}
  \bibinfo{person}{Liang Wang}.} \bibinfo{year}{2022}\natexlab{b}.
\newblock \showarticletitle{Evidence-aware fake news detection with graph
  neural networks}. In \bibinfo{booktitle}{\emph{TheWebConf}}.
  \bibinfo{pages}{2501--2510}.
\newblock


\bibitem[Xu et~al\mbox{.}(2012)]%
        {xu2012modeling}
\bibfield{author}{\bibinfo{person}{Zhiheng Xu}, \bibinfo{person}{Yang Zhang},
  \bibinfo{person}{Yao Wu}, {and} \bibinfo{person}{Qing Yang}.}
  \bibinfo{year}{2012}\natexlab{}.
\newblock \showarticletitle{Modeling user posting behavior on social media}. In
  \bibinfo{booktitle}{\emph{SIGIR}}. \bibinfo{pages}{545--554}.
\newblock


\bibitem[Yang et~al\mbox{.}(2022a)]%
        {yang2022weakly}
\bibfield{author}{\bibinfo{person}{Ruichao Yang}, \bibinfo{person}{Jing Ma},
  \bibinfo{person}{Hongzhan Lin}, {and} \bibinfo{person}{Wei Gao}.}
  \bibinfo{year}{2022}\natexlab{a}.
\newblock \showarticletitle{A weakly supervised propagation model for rumor
  verification and stance detection with multiple instance learning}. In
  \bibinfo{booktitle}{\emph{SIGIR}}. \bibinfo{pages}{1761--1772}.
\newblock


\bibitem[Yang et~al\mbox{.}(2022b)]%
        {yang2022reinforcement}
\bibfield{author}{\bibinfo{person}{Ruichao Yang}, \bibinfo{person}{Xiting
  Wang}, \bibinfo{person}{Yiqiao Jin}, \bibinfo{person}{Chaozhuo Li},
  \bibinfo{person}{Jianxun Lian}, {and} \bibinfo{person}{Xing Xie}.}
  \bibinfo{year}{2022}\natexlab{b}.
\newblock \showarticletitle{Reinforcement subgraph reasoning for fake news
  detection}. In \bibinfo{booktitle}{\emph{KDD}}. \bibinfo{pages}{2253--2262}.
\newblock


\bibitem[Yin et~al\mbox{.}(2019)]%
        {yin2019social}
\bibfield{author}{\bibinfo{person}{Hongzhi Yin}, \bibinfo{person}{Qinyong
  Wang}, \bibinfo{person}{Kai Zheng}, \bibinfo{person}{Zhixu Li},
  \bibinfo{person}{Jiali Yang}, {and} \bibinfo{person}{Xiaofang Zhou}.}
  \bibinfo{year}{2019}\natexlab{}.
\newblock \showarticletitle{Social influence-based group representation
  learning for group recommendation}. In \bibinfo{booktitle}{\emph{ICDE}}.
  \bibinfo{pages}{566--577}.
\newblock


\bibitem[Yoo et~al\mbox{.}(2023)]%
        {YooLSK23}
\bibfield{author}{\bibinfo{person}{Hyunsik Yoo}, \bibinfo{person}{Yeon{-}Chang
  Lee}, \bibinfo{person}{Kijung Shin}, {and} \bibinfo{person}{Sang{-}Wook
  Kim}.} \bibinfo{year}{2023}\natexlab{}.
\newblock \showarticletitle{Disentangling Degree-related Biases and Interest
  for Out-of-Distribution Generalized Directed Network Embedding}. In
  \bibinfo{booktitle}{\emph{TheWebConf}}. \bibinfo{pages}{231--239}.
\newblock


\bibitem[Zhang et~al\mbox{.}(2021)]%
        {zhang2021mining}
\bibfield{author}{\bibinfo{person}{Jinghao Zhang}, \bibinfo{person}{Yanqiao
  Zhu}, \bibinfo{person}{Qiang Liu}, \bibinfo{person}{Shu Wu},
  \bibinfo{person}{Shuhui Wang}, {and} \bibinfo{person}{Liang Wang}.}
  \bibinfo{year}{2021}\natexlab{}.
\newblock \showarticletitle{Mining Latent Structures for Multimedia
  Recommendation}. In \bibinfo{booktitle}{\emph{ACM MM}}.
  \bibinfo{pages}{3872--3880}.
\newblock


\bibitem[Zhang et~al\mbox{.}(2023a)]%
        {zhang2023efficiently}
\bibfield{author}{\bibinfo{person}{Peiyan Zhang}, \bibinfo{person}{Jiayan Guo},
  \bibinfo{person}{Chaozhuo Li}, \bibinfo{person}{Yueqi Xie},
  \bibinfo{person}{Jae~Boum Kim}, \bibinfo{person}{Yan Zhang},
  \bibinfo{person}{Xing Xie}, \bibinfo{person}{Haohan Wang}, {and}
  \bibinfo{person}{Sunghun Kim}.} \bibinfo{year}{2023}\natexlab{a}.
\newblock \showarticletitle{Efficiently leveraging multi-level user intent for
  session-based recommendation via atten-mixer network}. In
  \bibinfo{booktitle}{\emph{WSDM}}. \bibinfo{pages}{168--176}.
\newblock


\bibitem[Zhang and Kim(2023)]%
        {zhang2023survey}
\bibfield{author}{\bibinfo{person}{Peiyan Zhang} {and} \bibinfo{person}{Sunghun
  Kim}.} \bibinfo{year}{2023}\natexlab{}.
\newblock \showarticletitle{A Survey on Incremental Update for Neural
  Recommender Systems}.
\newblock \bibinfo{journal}{\emph{arXiv:2303.02851}} (\bibinfo{year}{2023}).
\newblock


\bibitem[Zhang et~al\mbox{.}(2023b)]%
        {zhang2023continual}
\bibfield{author}{\bibinfo{person}{Peiyan Zhang}, \bibinfo{person}{Yuchen Yan},
  \bibinfo{person}{Chaozhuo Li}, \bibinfo{person}{Senzhang Wang},
  \bibinfo{person}{Xing Xie}, \bibinfo{person}{Guojie Song}, {and}
  \bibinfo{person}{Sunghun Kim}.} \bibinfo{year}{2023}\natexlab{b}.
\newblock \showarticletitle{Continual Learning on Dynamic Graphs via Parameter
  Isolation}.
\newblock \bibinfo{journal}{\emph{arXiv:2305.13825}} (\bibinfo{year}{2023}).
\newblock


\bibitem[Zhou et~al\mbox{.}(2020)]%
        {zhou2020variational}
\bibfield{author}{\bibinfo{person}{Fan Zhou}, \bibinfo{person}{Xovee Xu},
  \bibinfo{person}{Kunpeng Zhang}, \bibinfo{person}{Goce Trajcevski}, {and}
  \bibinfo{person}{Ting Zhong}.} \bibinfo{year}{2020}\natexlab{}.
\newblock \showarticletitle{Variational information diffusion for probabilistic
  cascades prediction}. In \bibinfo{booktitle}{\emph{IEEE INFOCOM}}.
  \bibinfo{pages}{1618--1627}.
\newblock


\bibitem[Zhu et~al\mbox{.}(2021)]%
        {zhu2021deep}
\bibfield{author}{\bibinfo{person}{Yanqiao Zhu}, \bibinfo{person}{Weizhi Xu},
  \bibinfo{person}{Jinghao Zhang}, \bibinfo{person}{Qiang Liu},
  \bibinfo{person}{Shu Wu}, {and} \bibinfo{person}{Liang Wang}.}
  \bibinfo{year}{2021}\natexlab{}.
\newblock \showarticletitle{Deep graph structure learning for robust
  representations: A survey}.
\newblock \bibinfo{journal}{\emph{arXiv:2103.03036}} (\bibinfo{year}{2021}).
\newblock


\bibitem[Zhu et~al\mbox{.}(2022)]%
        {zhu2022structure}
\bibfield{author}{\bibinfo{person}{Yanqiao Zhu}, \bibinfo{person}{Yichen Xu},
  \bibinfo{person}{Hejie Cui}, \bibinfo{person}{Carl Yang},
  \bibinfo{person}{Qiang Liu}, {and} \bibinfo{person}{Shu Wu}.}
  \bibinfo{year}{2022}\natexlab{}.
\newblock \showarticletitle{Structure-enhanced heterogeneous graph contrastive
  learning}. In \bibinfo{booktitle}{\emph{SDM}}. \bibinfo{pages}{82--90}.
\newblock


\bibitem[Ziems et~al\mbox{.}(2020)]%
        {Ziems2020RacismIA}
\bibfield{author}{\bibinfo{person}{Caleb Ziems}, \bibinfo{person}{Bing He},
  \bibinfo{person}{Sandeep Soni}, {and} \bibinfo{person}{Srijan Kumar}.}
  \bibinfo{year}{2020}\natexlab{}.
\newblock \showarticletitle{Racism is a virus: anti-asian hate and
  counterspeech in social media during the COVID-19 crisis}.
\newblock \bibinfo{journal}{\emph{ASONAM}} (\bibinfo{year}{2020}).
\newblock


\end{thebibliography}

\appendix



\section{Discussion}

\begin{table*}
\centering
\caption{Performances on the \textsf{Small} dataset for warm-start videos}
\label{tab:warmSmall}
\vspace{-0.2cm}
\textbf{(a) Popular Subreddits}
\setlength{\tabcolsep}{4pt}
\begin{tabular}{c|cccccccc|c}
\toprule
& MF & NGCF & LightGCN & SVD-GCN & TiSASRec & TGAT & TGN & INPAC & Impr. \\
\midrule
NDCG@5 & 35.92 $\pm$ 0.30 & 36.39 $\pm$ 0.56 & 37.95 $\pm$ 0.53 & 38.89 $\pm$ 0.33 & 39.17 $\pm$ 0.50 & 39.22 $\pm$ 0.46 & \ul{40.33} $\pm$ 0.22 & \textbf{43.81 $\pm$ 0.29} & 8.6\% \\
Rec@5 & 52.21 $\pm$ 0.48 & 53.01 $\pm$ 0.41 & 55.05 $\pm$ 0.52 & 56.50 $\pm$ 0.37 & 56.34 $\pm$ 0.50 & 56.79 $\pm$ 0.17 & \ul{57.33} $\pm$ 0.30 & \textbf{60.89 $\pm$ 0.40} & 6.2\% \\
NDCG@10 & 40.25 $\pm$ 0.79 & 40.91 $\pm$ 0.25 & 41.77 $\pm$ 0.36 & 42.45 $\pm$ 0.24 & 42.95 $\pm$ 0.39 & 42.65 $\pm$ 0.16 & \ul{43.27} $\pm$ 0.25 & \textbf{46.11 $\pm$ 0.26} & 6.6\% \\
Rec@10 & 65.83 $\pm$ 0.52 & 66.30 $\pm$ 0.66 & 68.21 $\pm$ 0.61 & 68.48 $\pm$ 0.37 & 68.39 $\pm$ 0.33 & 68.53 $\pm$ 0.31 & \ul{68.60} $\pm$ 0.20 & \textbf{70.31 $\pm$ 0.59} & 2.5\% \\
MRR & 33.47 $\pm$ 0.61 & 34.12 $\pm$ 0.31 & 34.71 $\pm$ 0.38 & 36.22 $\pm$ 0.29 & 36.55 $\pm$ 0.46 & 36.75 $\pm$ 0.29 & \ul{37.45} $\pm$ 0.30 & \textbf{40.24 $\pm$ 0.44} & 7.5\% \\
\bottomrule
\end{tabular}

\textbf{(b) Non-Popular Subreddits}
\label{tab:coldSmall}
\begin{tabular}{c|cccccccc|c}
\toprule
NDCG@5 & 7.71 $\pm$ 0.31 & 8.14 $\pm$ 0.07 & 8.58 $\pm$ 0.23 & 9.50 $\pm$ 0.17 & 9.65 $\pm$ 0.22 & 9.64 $\pm$ 0.17 & \ul{9.87} $\pm$ 0.03 & \textbf{11.14} $\pm$ 0.18 & 12.9\% \\
Rec@5 & 12.00 $\pm$ 0.33 & 12.52 $\pm$ 0.18 & 13.16 $\pm$ 0.22 & 14.17 $\pm$ 0.11 & 14.38 $\pm$ 0.23 & 14.47 $\pm$ 0.18 & \ul{14.58} $\pm$ 0.12 & \textbf{15.31} $\pm$ 0.17 & 5.0\% \\
NDCG@10 & 9.93 $\pm$ 1.08 & 10.09 $\pm$ 0.13 & 11.96 $\pm$ 0.21 & 12.05 $\pm$ 0.13 & 12.37 $\pm$ 0.31 & 12.61 $\pm$ 0.32 & \ul{12.98} $\pm$ 0.08 & \textbf{14.25} $\pm$ 0.31 & 9.8\% \\
Rec@10 & 19.61 $\pm$ 0.46 & 18.21 $\pm$ 0.11 & 22.56 $\pm$ 0.31 & 23.16 $\pm$ 0.32 & 23.51 $\pm$ 0.31 & 23.27 $\pm$ 0.20 & \ul{23.63} $\pm$ 0.10 & \textbf{25.28} $\pm$ 0.44 & 7.0\% \\
MRR & 8.08 $\pm$ 0.24 & 9.09 $\pm$ 0.35 & 9.74 $\pm$ 0.20 & 10.20 $\pm$ 0.17 & 10.60 $\pm$ 0.15 & 11.11 $\pm$ 0.20 & \ul{11.57} $\pm$ 0.16 & \textbf{13.75} $\pm$ 0.21 & 18.8\% \\
\bottomrule
\end{tabular}
\end{table*}

\subsection{Difference between \problem and Recommendation Problems}


\subsubsection{Distinct underlying dynamics} 
In recommendation problems, the focus is on user behavior, as it largely reflects their interests, making it crucial to model user preferences accurately for precise recommendations. Group recommendation models~\cite{yin2019social} suggest items based on the combined preferences of users in a group, whereas sequential recommendation~\cite{wu2019session, 10.1145/3543507.3583503} models concentrate on individual users' preferences and the extent to which item attributes align with those preferences.

On the other hand, the \problem problem encompasses a combination of factors that influence a user's decision to post a video within a community, where different users can share the video on different communities. An information-sharing event within a community is subject to factors such as user interests, community characteristics, and the relationship between the community and the information being shared. For example, a piece of information can be posted in some online community due to the following reasons:

\begin{itemize}[leftmargin=*]
    \item Community members find the information valuable and wish to share it with other, driven by internal factors such as interest or altruism; 
    \item Some users who originally do not belong to the community want to promote their product or service to a wider audience; 
    \item Users with malicious intent seek to spread false or misleading information
\end{itemize}

\subsubsection{The User Behaviors to be Modeled are Different}
The goal of the proposed \problem problem requires simultaneous understanding of multiple users’ behavior. One video can be shared by different users on different communities with completely different motivations. For example, a video $v_1$ can be shared in community $s_1$ by user $u_1$ with positive intent (\eg, promoting the video) while by another user $u_2$ in community $s_2$ with negative intent (\eg, criticizing the video). Yet, the goal is still to predict the next community  on which the video will appear. 

\subsubsection{The Goals of the Problems are Different}

The primary objective of the \problem problem is to model information flow across online communities rather than creating a recommender system. Although our proposed INPAC approach can be adapted for sequential recommendation, its primary focus is on capturing the complex interactions between users, communities, and information. Our experimental results (Tables 4-5) demonstrate that existing recommendation models were not designed to address the \problem problem and have inherent limitations when applied to it.

\subsubsection{The Datasets are not Directly Transferable}

As the first three points suggest, existing recommender system datasets, such as LastFM\footnote{\url{http://millionsongdataset.com/lastfm/}}, MovieLens\footnote{\url{https://grouplens.org/datasets/movielens/}}, and Goodreads\footnote{\url{https://sites.google.com/eng.ucsd.edu/ucsdbookgraph/home}}, are not directly applicable to solving the \problem problem, as they lack information about clearly defined online communities and the sharing of information across those online communities. This discrepancy highlights the need for distinct datasets that capture the complex dynamics specific to the \problem problem.

In summary, although there may be some overlap between the methods used in recommender systems and the \problem problem, they are fundamentally different problems that require distinct approaches to model the unique interactions between users, communities, and information sharing.

\begin{table*}
\centering
\caption{Performances on the \textsf{Small} dataset for cold-start videos. }
\setlength{\tabcolsep}{4pt}
\label{tab:coldSmall}
\vspace{-0.2cm}
\textbf{(a) Popular Subreddits}
\begin{tabular}{c|cccccccc|c}
\toprule
& MF & NGCF & LightGCN & SVD-GCN & TiSASRec & TGAT & TGN & INPAC & Impr. \\
    \midrule
    NDCG@5 & 34.57 $\pm$ 1.56 & 36.46 $\pm$ 0.17 & 39.24 $\pm$ 0.97 & 40.27 $\pm$ 1.29 & 41.33 $\pm$ 0.94 & {42.45} $\pm$ 1.16 & \ul{42.66} $\pm$ 1.16 & \textbf{46.44} $\pm$ 1.28 & 8.9\% \\
    Rec@5 & 58.19 $\pm$ 1.85 & 58.51 $\pm$ 0.51 & 57.58 $\pm$ 0.17 & 60.60 $\pm$ 0.95 & 64.46 $\pm$ 0.90 & {67.07} $\pm$ 1.30 & \ul{67.51} $\pm$ 1.27 & \textbf{71.53} $\pm$ 1.17 & 6.0\% \\
    NDCG@10 & 41.02 $\pm$ 0.82 & 43.63 $\pm$ 0.16 & 43.75 $\pm$ 0.99 & 44.74 $\pm$ 0.64 & 46.11 $\pm$ 1.10 & {47.77} $\pm$ 1.17 & \ul{47.90} $\pm$ 1.08 & \textbf{50.81} $\pm$ 1.33 & 6.1\% \\
    Rec@10 & 84.98 $\pm$ 0.86 & 83.64 $\pm$ 0.41 & 85.53 $\pm$ 0.26 & 87.52 $\pm$ 0.89 & 88.13 $\pm$ 0.99 & {88.13} $\pm$ 1.14 & \ul{88.19} $\pm$ 1.13 & \textbf{91.17} $\pm$ 1.41 & 3.4\% \\
    MRR & 29.65 $\pm$ 0.76 & 31.95 $\pm$ 0.46 & 31.93 $\pm$ 0.76 & 32.56 $\pm$ 0.76 & 36.55 $\pm$ 1.27 & {36.64} $\pm$ 1.24 & \ul{36.89} $\pm$ 1.14 & \textbf{38.44} $\pm$ 1.27 & 4.2\% \\
\bottomrule
\end{tabular}

\textbf{(b) Non-Popular Subreddits}
\begin{tabular}{c|cccccccc|c}
\toprule
    NDCG@5 & 7.23 $\pm$ 1.18 & 8.16 $\pm$ 0.38 & 8.04 $\pm$ 1.09 & 7.86 $\pm$ 0.95 & 8.45 $\pm$ 1.36 & 8.39 $\pm$ 1.15 & \ul{8.83} $\pm$ 1.12 & \textbf{10.05} $\pm$ 1.20 & 13.8\% \\
    Rec@5 & 10.53 $\pm$ 1.02 & 12.30 $\pm$ 0.50 & 11.72 $\pm$ 0.38 & 13.41 $\pm$ 0.80 & 13.23 $\pm$ 1.28 & 13.92 $\pm$ 1.29 & \ul{14.40} $\pm$ 1.14 & \textbf{15.23} $\pm$ 0.81 & 5.8\% \\
    NDCG@10 & 11.59 $\pm$ 1.76 & 11.12 $\pm$ 0.30 & 10.55 $\pm$ 0.90 & 11.27 $\pm$ 1.06 & 10.13 $\pm$ 0.32 & 11.50 $\pm$ 1.18 & \ul{11.62} $\pm$ 1.16 & \textbf{12.75} $\pm$ 0.21 & 9.7\% \\
    Rec@10 & 20.40 $\pm$ 0.86 & 23.07 $\pm$ 0.34 & 22.78 $\pm$ 0.76 & 22.72 $\pm$ 1.42 & 23.38 $\pm$ 1.32 & 23.71 $\pm$ 1.21 & \ul{23.90} $\pm$ 1.09 & \textbf{25.09} $\pm$ 1.20 & 5.0\% \\
    MRR & 8.61 $\pm$ 1.34 & 9.78 $\pm$ 0.24 & 10.98 $\pm$ 0.38 & 10.00 $\pm$ 0.43 & 9.30 $\pm$ 0.16 & 10.68 $\pm$ 1.21 & \ul{10.97} $\pm$ 1.18 & \textbf{11.83} $\pm$ 1.12 & 7.8\% \\
\bottomrule
\end{tabular}
\end{table*}
\subsection{Extension to Other Types of Features}

Our proposed framework can be extended to handle other complex types of information, such as images and audio. 
We outline the simple modifications required to accommodate these data formats. Specifically, in Section~\ref{sec:video}, we can add the appropriate encoders for image or audio instead of using the encoder for video content. Below are the potential encoders to use image and audio content

\textbf{Image}: To handle images, we can incorporate a variety of image encoders, such as CNN~\cite{lecun1995convolutional}, ResNet~\cite{he2016deep}, or Vision Transformer~\cite{dosovitskiyimage, xu2022tervit}, which will convert each input image into a $D$-dimensional feature vector. This vector can then be fed into our current model architecture as an input for community prediction.

\textbf{Audio} To accommodate audio data, there are several options for encoding audio into a $D$-dimensional representation, including MFCC-based models~\cite{logan2000mel}, 
LSTM~\cite{graves2012long}, or Transformer-based models such as Conformer~\cite{gulati2020conformer}. The choice of encoder would depend on the specific characteristics of the audio data, the acceptable level of computational costs, and the desired level of representation.

\subsection{Rationale Behind Using Reddit Data}

The rationale for focusing on YouTube videos on Reddit is:
\begin{itemize}[leftmargin=*]
    \item \textbf{Reddit}: We chose to study Reddit because Reddit is one of the largest global social platforms. It is ranked among the top 10 visited websites worldwide~\cite{wiki-most-visited-website}. Most importantly, unlike many other social platforms, Reddit users form clearly defined community structures. These communities, also known as subreddits, are typically centered around specific topics or interests, such as music, politics, science, or gaming. The community-centric nature of Reddit makes it easy to analyze user group behavior and identify patterns of information sharing across different communities.
    \item \textbf{Cross-posting of YouTube videos on Reddit}: YouTube videos have previously been shown to be a major means of spreading misinformation on other platforms including Reddit~\cite{micallef2020role}. YouTube is the second most popular platform in the world with 2.51 billion monthly active users~\cite{statista-social-networks}, and is one of the most popular ways for users to consume online information.
    \item \textbf{Rich semantic information}: YouTube videos contain a wealth of textual and visual information that can help us develop a more comprehensive understanding of the content and its potential for spreading misinformation. This multimodal nature allows us to extract features from both visual and textual data.
    \item \textbf{Traceability of sharing patterns}: The sharing of YouTube videos on Reddit can be easily traced, enabling us to study the dissemination of misinformation across communities. In contrast, other types of information such as quotes and online memes ~\cite{ling2021dissecting, qu2022evolution, leskovec2009meme}, can be more difficult to track due to the evolution and modification of their content ~\cite{qu2022evolution, leskovec2009meme}.

\end{itemize}

\section{Evaluation}\label{app:evaluation}

\subsection{Performances on the \textsf{Small} Dataset}

Tables~\ref{tab:warmSmall} and \ref{tab:coldSmall} present the results on the \textsf{Small} dataset for warm-start and cold-start videos, respectively. Remarkably, our \model model consistently surpasses all baseline methods with statistically significant improvements.



\end{document}